\documentclass[aps,pra,reprint,superscriptaddress,nofootinbib,floatfix,fleqn]{revtex4-2}

\usepackage[utf8]{inputenc}
\usepackage[T1]{fontenc}
\usepackage{csquotes}
\usepackage[greek,main=english]{babel}

\usepackage[a4paper,margin=2cm]{geometry}

\usepackage[font=normalsize,caption=false]{subfig}

\usepackage{anyfontsize}
\usepackage[varvw]{newtx}
\DeclareFontFamilySubstitution{LGR}{ntxtlf}{Tempora-TLF}
\usepackage{microtype}
\usepackage{tabularray}
\UseTblrLibrary{booktabs}
\allowdisplaybreaks

\usepackage[dvipsnames]{xcolor}
\usepackage[hidelinks,colorlinks=true,allcolors=Blue]{hyperref}

\usepackage{mathtools}

\usepackage{amsfonts,mathrsfs,dutchcal,textgreek}
\let\bs\boldsymbol

\usepackage{tensor}
\let\t\tensor
\usepackage{slashed}

\usepackage{braket}

\def\e{\mathrm e}	
\def\i{\mathrm i}	
\def\d{\mathrm d}	
\def\dd{\,\d}		
\def\p{\partial}	
\def\I{\textsc{i}}
\def\II{\textsc{ii}}

\def\half{\tfrac{1}{2}}

\def\third{\tfrac{1}{3}}

\def\quarter{\tfrac{1}{4}}

\def\Mfd{\mathscr M}

\def\Momentum{P}

\def\Velocity{V}
\def\velocity{\varv}

\def\Metric{g}

\def\Riemann{R}

\def\metric{l}
\def\riemann{P}
\newcommand{\Christoffel}[3]{\left\{\begin{matrix}#1\\#2#3\end{matrix}\right\}}
\newcommand{\christoffel}[3]{\{\substack{#1\\#2#3}\}}

\def\lapse{\zeta}
\def\llapse{\varphi}
\def\acc{a}
\def\shift{\xi}
\def\twist{\Xi}

\def\Killing{\mathcal K}
\def\Lorentz{\varGamma}

\def\congr{\upsilon}

\def\length{l}

\def\Eikonal{\Psi}
\def\eikonal{\psi}
\def\WaveC{K}
\def\waveC{k}
\def\WaveV{\bar K}

\def\waveN{\hat n}

\def\MetricO{\bar g}
\def\DelO{\bar \nabla}

\def\gravA{\varg} 
\def\Tret{\mathcal T}

\def\ADM{\textsc{adm}}
\def\BMV{\textsc{bmv}}
\def\COW{\textsc{cow}}
\def\EM{\textsc{em}}
\def\ERC{\textsc{erc}}
\def\etal{\textit{et al}}
\def\GRAVITES{\textsc{gravites}}
\def\GW{\textsc{gw}}
\def\IQOQI{\textsc{iqoqi}}
\def\MPD{\textsc{mpd}}
\def\PPN{\textsc{ppn}}
\def\QFT{\textsc{qft}}
\def\TURIS{\textsc{turis}}
\def\VDSP{\textsc{vdsp}}
\def\WKB{\textsc{wkb}}
\def\FSL{\textsc{fsl}}

\usepackage{siunitx}
\DeclareSIUnit{\foot}{ft}

\usepackage[capitalize]{cleveref}
\usepackage{orcidlink}
\usepackage{aas_macros}

\usepackage{enumitem}
\setlist[enumerate]{
	label=(\roman*),
	leftmargin=0pt,
	itemindent=\dimexpr\labelwidth+\labelsep\relax
}

\hypersetup{
	pdftitle={Quantum interferometry in external gravitational fields},
	pdfsubject={},
	pdfauthor={Thomas B. Mieling, Thomas Morling, Christopher Hilweg, and Philip Walther}
}

\begin{document}
\title{Quantum interferometry in external gravitational fields}
\author{Thomas B. Mieling~\orcidlink{0000-0002-6905-0183}}
\email{thomas.mieling@univie.ac.at}
\affiliation{University of Vienna, Faculty of Physics and Research Network \TURIS, Boltzmanngasse~5, 1090 Vienna, Austria}
\author{Thomas Morling~\orcidlink{0009-0009-3373-5349}}
\affiliation{University of Vienna, Faculty of Physics, Vienna Doctoral School in Physics~(\VDSP) and Research Network \TURIS, Boltzmanngasse~5, 1090 Vienna, Austria}
\author{Christopher Hilweg~\orcidlink{0000-0003-0213-1234}}
\affiliation{University of Vienna, Faculty of Physics and Research Network \TURIS, Boltzmanngasse~5, 1090 Vienna, Austria}
\author{Philip Walther~\orcidlink{0000-0002-4964-817X}}
\affiliation{University of Vienna, Faculty of Physics and Research Network \TURIS, Boltzmanngasse~5, 1090 Vienna, Austria}
\affiliation{Institute for Quantum Optics and Quantum Information (\IQOQI) Vienna, Austrian Academy of Sciences, Boltzmanngasse 3, 1090 Vienna, Austria}
\begin{abstract}
	Current models of quantum interference experiments in external gravitational fields lack a common framework: while matter-wave interferometers are commonly described using the Schrödinger equation with a Newtonian potential, gravitational effects in quantum optics are modeled using either post-Newtonian metrics or highly symmetric exact solutions to Einstein’s field equations such as those of Schwarzschild and Kerr.
	To coherently describe both kinds of experiments, this paper develops a unified framework for modeling quantum interferometers in general stationary space-times. This model provides a rigorous description and coherent interpretation of the effects of classical gravity on quantum probes.
\end{abstract}
\maketitle

\begingroup%
\renewcommand\thefootnote{}%
\footnote{\newline\newline
Based on the article published by APS in \textit{Physical Review A},
DOI:~\href{https://doi.org/10.1103/8dsx-jt6b}{10.1103/8dsx-jt6b}, under the terms of the
\href{https://creativecommons.org/licenses/by/4.0/}{CC~BY~4.0 license}. Further distribution of this work must maintain attribution
to the author(s) and the published article’s title, journal citation, and DOI.
}%
\addtocounter{footnote}{-1}%
\endgroup%
\vspace*{-3\baselineskip}

\section{Introduction}

A central question in current theoretical physics is how quantum theory and general relativity can be unified into a consistent overarching framework. While quantum field theory (\QFT) provides a generally covariant extension of quantum mechanics to flat and curved space-times, this framework treats gravity as a classical background field.
There are multiple approaches to formulating a quantum theory of gravity, such as loop quantum gravity \cite{2008LRR....11....5R}, string theory \cite{2024SenZwiebach}, and others \cite{2020CQGra..37a3002L,2006LRR.....9....5N,2019LRR....22....5S,1973PhR.....6..241P}, but also proposals for theories in which the gravitational field remains classical \cite{2014FoPh...44..557P,2023PhRvX..13d1040O}.
The lack of consensus on the accuracy of these models is strongly tied to the absence of direct experimental tests: The energies required to study quantum gravity effects are beyond current technological capabilities and could remain so for the coming decades \cite{2019CQGra..36c4001C,2025RvMP...97a5003B}.

Nevertheless, innovative experiments and proposals have been put forward to investigate gravitational effects in quantum physics, utilizing both massive and massless quantum systems under low-energy conditions in weak gravitational fields.
In particular, the so-called \BMV\ experiments \cite{2017PhRvL.119x0401B,2017PhRvL.119x0402M} aim at detecting the emergence of entanglement between two quantum particles that interact with each other solely through gravity. There is an ongoing debate about what could be inferred from the observation of such gravitationally mediated entanglement \cite{2018JPhA...51h5303H,2019PhLB..792...64C,2021PhRvD.104l6030K,2022Quant...6..779G,2023PhRvD.108j1702M,2025RvMP...97a5003B}. A complementary approach is to study the behavior of quantum systems in classical gravitational fields and to explore the gravitationally influenced dynamics of entangled quantum states. Unlike \BMV\ experiments, this approach does not seek evidence for the quantum nature of gravity but tests the validity of various formulations of the equivalence principle and related phenomena such as the gravitational redshift with quantum resources.

The first experimental investigations in this direction focused on the behavior of massive quantum particles in Earth’s gravitational field \cite{1975PhRvL..34.1472C,1988PhyBC.151...22W,1997PhRvA..56.1767L}. These groundbreaking experiments showed that neutrons in spatial superpositions exhibit measurable phase shifts due to gravitational acceleration. Although the results can be interpreted in general relativistic terms \cite{2000AmJPh..68..404V}, they are also adequately explained using the Schrödinger equation with a Newtonian gravitational potential \cite{1998PhRvA..57.1260M,2008PrPNP..60....1A,2023_Huggett,2025RvMP...97a5003B}. Subsequent experiments with matter-wave interferometers further explored the response of matter-waves to gravitational fields \cite{1991PhRvL..67..181K,1992ApPhB..54..321K,1998PhRvL..81..971S,2013NJPh...15b3009A,2019Sci...366..745X,2024Natur.631..515P}, even revealing phase shifts induced by gravity gradients \cite{2022Sci...375..226O}.
Photonic experiments, on the other hand, could investigate gravitational effects on quantum systems in hitherto unexplored aspects by contrasting matter-wave interferometers with similar setups in which the probes propagate at relativistic speeds \cite{1979GReGr..11..391S,1983PhRvL..51..378T,2012CQGra..29v4010Z,2017NJPh...19c3028H,2022PhRvA.106f3511M}, building large-scale interferometers using satellites \cite{2022EPJQT...9...25M,2012CQGra..29v4011R,2024PhRvL.133b0201W}, and using entangled multi-photon states \cite{2022PhRvA.106c1701M,2024Quant...8.1273B}.

Current theoretical descriptions of massive and massless quantum probes do not share a common framework.
Specifically, the literature on gravitational photon interferometers predominantly uses Einsteinian models of gravity, see, e.g., Refs.~\cite{2012CQGra..29v4010Z,2017NJPh...19c3028H,2022PhRvA.106f3511M,2022PhRvD.105j5016B}. In the literature on gravitational matter-wave interferometers, on the other hand, it is not uncommon to describe the gravitational field using Newtonian methods even when discussing Einsteinian concepts such as the equivalence principle, proper time, or space-time curvature, see, e.g., Refs.~\cite{2017PhRvL.118r3602A,2022Sci...375..226O,2024PhRvD.109f4073C}.
This discrepancy illustrates the ongoing challenge of finding adequate descriptions of gravitational effects in quantum systems even if quantum gravity effects are neglected.

The aim of this paper is to provide a consistent and general method for describing quantum interference in external gravitational fields.
This is based on the Einsteinian description of gravity using Lorentzian geometry, but does not require Einstein’s field equations to hold. In our model, the external gravitational field is assumed to be stationary, but it need not be static nor weak.
Despite this level of generality, the resulting equations are not significantly more complex than in Newtonian models, but offer greater clarity in their physical interpretation.

This paper is organized as follows. After introducing the general mathematical framework for describing experiments in stationary space-times in \cref{s:geometry} we provide general expressions for the phase evolution in interferometric setups in \cref{s:phase evolution}. The conversion from phase differences to detection probabilities for both bosonic and fermionic particles is provided in \cref{s:quantum probabilities}. Our framework is then applied to a variety of optical and matter-wave experiments in \cref{s:experiments:light,s:experiments:matter}, respectively. Throughout this paper, Latin indices $i$, $j$, $\ldots$ range from $1$ to $3$ whereas Greek indices $\mu$, $\nu$, $\ldots$ range from $0$ to $3$, and indices occurring once as a subscript and once as a superscript are summed over (Einstein summation convention). The metric signature is $(-,+,+,+)$ and natural constants such as $c$, the speed of light in vacuum, and $\hslash$, the reduced Planck constant, are kept explicit.

\section{Geometry of stationary space-times}
\label{s:geometry}

This section provides a general framework for describing stationary space-times that forms the basis for the models of concrete experiments developed in the following sections.
The position taken here is that it is both physically more transparent and mathematically simpler to use a geometric and covariant formalism than to use perturbative weak-field expansions from the onset.
Although explicit expressions are presented in adapted coordinates, the geometric approach taken here helps to clarify interpretational issues arising from general covariance.

\paragraph{Time-translations}
The condition for a space-time $(\Mfd, \Metric)$ to be stationary can be formulated in terms of time-translation invariance.
Specifically, such an invariance can be described by a smooth action of the additive group of real numbers $\mathbf R$ on the manifold $\Mfd$, here denoted by $\Phi : \mathbf R \times \Mfd \to \Mfd$ with $(\chi, p) \mapsto \Phi_\chi(p)$, such that
(i) the orbits of $\Phi$ are timelike, and
(ii) all transformations $\Phi_\chi : \Mfd \to \Mfd$ are isometries of the metric $g$.
The infinitesimal generator of $\Phi$ is known as the associated Killing vector field $\Killing$ \cite[Sect.~6.2]{2019_Chrusciel} whose components, in local coordinates, are given by $\t\Killing{^\mu} = \d \t*\Phi{^\mu_\chi} / \d \chi |_{\chi = 0}$.

\begin{figure}[b]
	\centering
	\subfloat[]{
		\includegraphics[width=0.3\columnwidth,clip,trim={1cm 0cm 1cm 0cm}]{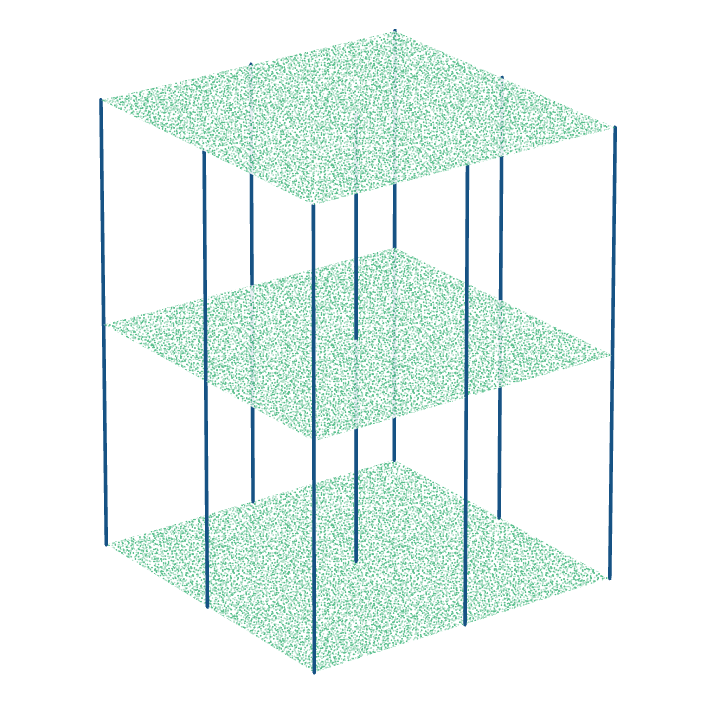}
	}
	\hfill
	\subfloat[]{
		\includegraphics[width=0.3\columnwidth,clip,trim={1cm 0cm 1cm 0cm}]{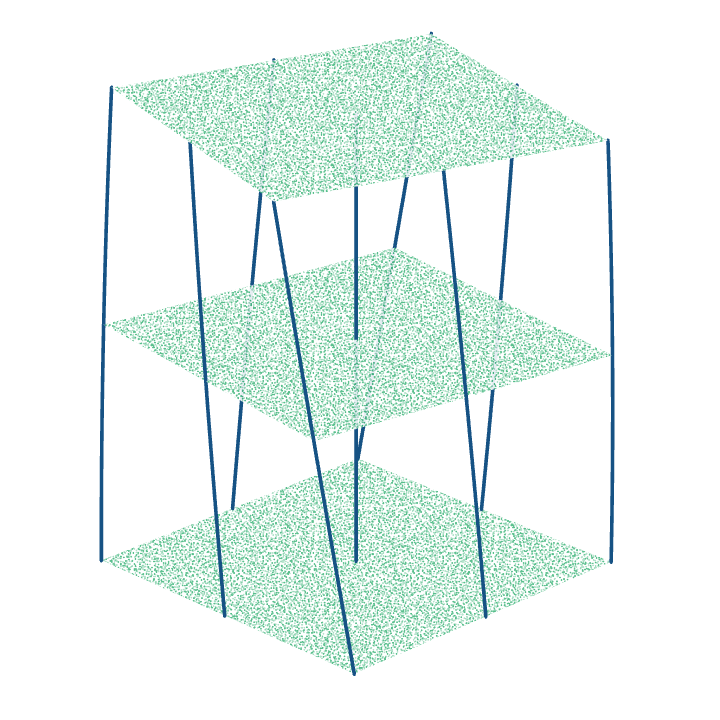}
	}
	\hfill
	\subfloat[]{
		\includegraphics[width=0.3\columnwidth,clip,trim={1cm 0cm 1cm 0cm}]{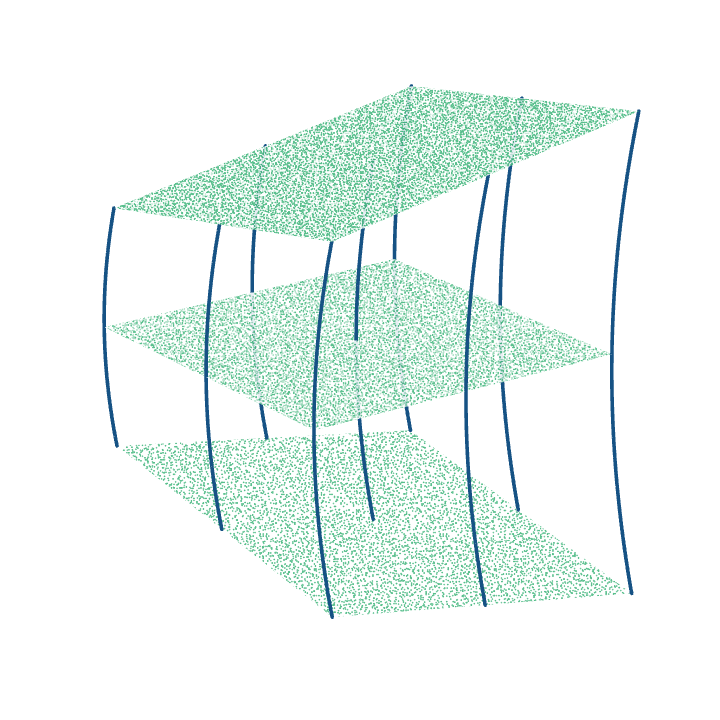}
	}
	\caption{
	Graphical illustration of different notions of isometric time-translations in flat space-time. The blue lines indicate the orbits of the group action $\Phi$.
	The central green surface is a plane of simultaneity for an inertial system; the surfaces above and below indicate how such a surface is deformed by different notions of time-translation: it is shifted in parallel under inertial motion (a), twisted under rotary motion (b), and tilted under hyperbolic motion (c).
	}
	\label{fig:time-translations}
\end{figure}

Examples in flat space-time include inertial motion, rigid rotation, and hyperbolic motion, see \cref{fig:time-translations} for an illustration.
Similarly, the exterior Schwarzschild geometry admits multiple notions of time-translations: they include translations in the temporal coordinate of the Schwarzschild chart, combinations of such translations with rotations in the angular coordinates of the same chart, as well as translations in the temporal coordinate of the Gullstrand–Painlevé chart \cite{2001AmJPh..69..476M}. Thus, depending on the degree of symmetry, a space-time may admit multiple notions of isometric time-translations (and, in general, no single such notion is “preferred”).

\paragraph{Lapse and acceleration}
The lapse function $\lapse = \sqrt{- \t\Metric{_\mu_\nu} \t\Killing{^\mu} \t\Killing{^\nu}} / c$ measures the rate of flow of proper-time $\tau$ along the orbits of $\Phi$ via $\d\tau = \lapse \d\chi$.
Hence, the four-velocity of these orbits is given by $\t\congr{^\mu} = \t\Killing{^\mu} / \lapse$.
In the following, the flow parameter $\chi$ will be assumed to have units of time, which implies that $\lapse$ is dimensionless.
This flow parameter can be rescaled according to $\chi \mapsto \chi' = \chi/\alpha$ (with constant $\alpha > 0$), which induces $\t\Killing{^\mu} \mapsto \t\Killing{^\prime^\mu} = \alpha \t\Killing{^\mu}$ and $\lapse \mapsto \lapse^\prime = \alpha \lapse$. As a consequence, the lapse can be set to unity at any given point.
However, it is not possible to set $\lapse = 1$ on open regions unless the orbits of $\Phi$ are inertial. This is because derivatives of $\lapse$ determine the four-acceleration via $\t\acc{_\mu} = c^2 (\t\p{_\mu} \lapse) / \lapse$.
Defining the logarithmic lapse as $\llapse = c^2 \ln \lapse$, this can be written as $\t\acc{_\mu} = \t\p{_\mu} \llapse$.

\paragraph{Synchronization and Lorentz factors}
Two events $\t p{_1}, \t p{_2} \in \Mfd$ are said to be simultaneous relative to $\Phi$ if there exists a connecting curve $\alpha \mapsto \t x{^\mu}(\alpha)$ satisfying $\t\Metric{_\mu_\nu} \t{\dot x}{^\mu} \t\Killing{^\nu} = 0$.
More generally, the one-form $\t\Metric{_\mu_\nu} \t\Killing{^\nu}$ does not only define a local notion of simultaneity, but also provides a measure of proper time ratios (that depends on the choice of $\Phi$).
Specifically, setting $\t\Lorentz{_\mu} = - \t\Metric{_\mu_\nu} \t\Killing{^\nu} / (c^2 \lapse)$, the Lorentzian gamma factor that relates the rate of proper time along integral curves of $\t\congr{^\mu} = \t\Killing{^\mu}/\lapse$ to that along any other timelike curve with four-velocity $\t u{^\mu}$ is given by $\gamma(\congr, u) = \t\Lorentz{_\mu} \t u{^\mu}$ [clearly, one has $\gamma(\congr, \congr) = \t\Lorentz{_\mu} \t\congr{^\mu} = 1$]. Based on the relative Lorentz factor $\gamma(\congr, u)$, the relative speed $V_{\operatorname{rel.}}(\congr, u)$ can be defined using the standard formula $\gamma(\congr, u) = 1/\sqrt{1 - V_{\operatorname{rel.}}(\congr, u)^2/c^2}$.

\paragraph{Shift and twist}
Locally, one can construct coordinates $(t, \t x{^i})$ such that $\Phi$ acts as $\Phi_\chi(t, \t x{^i}) = (t + \chi, \t x{^i})$, which is equivalent to $\t\Killing{^0} = 1$ and $\t\Killing{^i} = 0$ \cite[Sect.~1.4c]{2012_Frankel}. In any such coordinate system one thus has $\t\Killing{_0} = - c^2 \lapse^2$. The components $\t\Killing{_i}$, however, are not fully determined by the geometry of the problem and thus exhibit a gauge redundancy. Specifically, redefining the temporal coordinate as $t \mapsto t' = t + \theta$, where $\theta$ is a function of the “spatial coordinates” $\t x{^i}$ alone, the induced transformation reads $\t\Killing{_i} \mapsto \t*\Killing{^\prime_i} = \t\Killing{_i} - \t\Killing{_0} \t\p{_i} \theta$. Defining the dimensionless shift (relative to a given coordinate system) as $\t\shift{_i} = - c \t\Killing{_i} / \t\Killing{_0}$, this can be written as $\t\shift{_i} \mapsto \t*\shift{^\prime_i} = \t\shift{_i} + \t\p{_i} \lambda$, where $\lambda = c \theta$.
This means that $\t\shift{_i}$ (and hence also $\t\Killing{_i}$ and $\t\Metric{_0_i}$) can be chosen to vanish at any given point.
However, the twist $\t\twist{_i_j} = c \lapse \t{(\d\shift)}{_i_j} = c \lapse (\t\p{_i} \t\shift{_j} - \t\p{_j} \t\shift{_i})$, which is invariant under such transformations, provides an obstruction to setting these quantities to zero on open regions.
This tensor has a geometric interpretation: the condition $\t\twist{_i_j} = 0$ is equivalent to $\t\Killing{^\mu}$ being hypersurface-orthogonal \cite[App.~B]{1984_Wald}.
Hence, $\t\twist{_i_j}$ quantifies the obstruction to clock synchronization along closed spatial paths, which gives rise to the Sagnac effect.
Space-times admitting Killing fields with vanishing twist are said to be \emph{static}, but in generic \emph{stationary} space-times one has $\t\twist{_i_j} \neq 0$.

\paragraph{Spatial metric and curvature}
Since the lapse $\lapse$ and the shift $\t\shift{_i}$ determine the metric coefficients $\t\Metric{_0_\mu}$ in adapted coordinates, it remains to study the spatial components $\t\Metric{_i_j}$. These coefficients, however, transform non-trivially under the gauge transformations described above, $\t\Metric{_i_j} \mapsto \t*\Metric{^\prime_i_j} = \t\Metric{_i_j} - \lapse^2[ \t\p{_i}\lambda \; \t\p{_j}\lambda + \t\shift{_i}\t\p{_j}\lambda + \t\shift{_j}\t\p{_i}\lambda ]$, and thus do not admit an invariant geometric interpretation. Instead, a gauge-invariant version is given by $\t\metric{_i_j} = \t\Metric{_i_j} + \lapse^2 \t\shift{_i} \t\shift{_j}$. This “spatial metric” admits the fully covariant expression $\t\metric{_\mu_\nu} = \t\Metric{_\mu_\nu} + \t\Killing{_\mu} \t\Killing{_\nu} / (c \lapse)^2$, which shows that $\t\metric{_i_j}$ describes the metric $\t\Metric{_\mu_\nu}$ in the orthogonal complement of $\t\Killing{^\mu}$.
By a suitable choice of spatial coordinates $\t x{^i}$, one may set $\t\metric{_i_j}$ equal to the Kronecker delta $\t\delta{_i_j}$ at any given point or along any given spatial curve (using, for example, spatial Riemann coordinates or Fermi coordinates), but the extension to open neighborhoods is possible if and only if the spatial curvature tensor, here denoted by $\t\riemann{_i_j_k_l}$, vanishes \cite[Chap.~7]{2018_Lee}.

\paragraph{Metric decomposition}
The constructions described so far show that any stationary metric admits local adapted coordinates $(t, \t x{^i})$ such that the time-translation map $\Phi$ acts as $\Phi_\chi(t, \t x{^i}) = (t + \chi, \t x{^i})$ and the space-time metric takes the form
\begin{align}
	\label{eq:Metric 3+1}
	\Metric = - \lapse^2 ( c \d t - \t\shift{_i} \t{\d x}{^i} )^2 + \t\metric{_i_j} \t{\d x}{^i} \t{\d x}{^j}\,,
\end{align}
where the lapse $\lapse$, shift $\t\shift{_i}$, and spatial metric $\t\metric{_i_j}$ are independent of the temporal coordinate $t$.
The remaining coordinate freedom covers space-dependent translations in $t$ (inducing $\t\shift{_i} \mapsto \t\shift{_i} + \t\p{_i} \lambda$), and $t$-independent spatial diffeomorphisms (under which $\lapse$, $\t\shift{_i}$, and $\t\metric{_i_j}$ transform as covariant tensors of rank 0, 1, and 2, respectively).

\paragraph{Frames of reference}
While it is common practice to define stationary metrics by the condition of admitting coordinates with respect to which $\t\Metric{_\mu_\nu}$ can be written in the form \eqref{eq:Metric 3+1}, it is worth noting that such a description specifies not only the metric $\t\Metric{_\mu_\nu}$, but also a notion of time-translations (namely translations in the temporal coordinate $t$).
Indeed, if the isometry group of $(\Mfd, \Metric)$ has multiple dimensions, the same metric tensor may take different forms depending on the choice of Killing field with respect to which the metric decomposition is carried out.
For example, the expressions
\begin{subequations}
\begin{align}
	\label{eq:Minkowski:inertial}
	\Metric_\textsc{\,i} &= - (c \d t)^2 + \d x^2 + \d y^2 + \d z^2\,,
	\\
	\label{eq:Minkowski:rotation}
	\begin{split}
		\Metric_\textsc{r} &=
			- \left[ 1 - \frac{r^2 \omega^2}{c^2} \right]
				\left[ c \d t - \frac{(r \omega/c)\, r \d \phi}{1 - r^2 \omega^2/c^2}\right]^2
			\\&\hspace{5em}
			+ \d r^2 + \frac{r^2 \d \phi^2}{1 - r^2 \omega^2/c^2} + \d z^2\,,
	\end{split}
	\\
	\label{eq:Minkowski:hyperbolic}
	\Metric_\textsc{h} &= - \left[1 + \frac{a z}{c^2}\right]^2 (c \d t)^2 + \d x^2 + \d y^2 + \d z^2
\end{align}
\end{subequations}
all describe the flat Minkowski geometry (or parts thereof), but define different notions of time-translations whose orbits are depicted in \cref{fig:time-translations}. In this sense, \cref{eq:Minkowski:inertial,eq:Minkowski:rotation,eq:Minkowski:hyperbolic} can be interpreted as defining different frames of reference. The situation is similar in the (exterior) Schwarzschild and Kerr geometries where “standard” time-translations can be combined with rotations to produce further notions of time-translations.
The possibility of there being multiple stationary reference systems is important when describing experiments “at rest” in a gravitational field as this notion becomes ambiguous in such scenarios. In such a case, the system considered as being at rest defines one notion of time-translations and hence singles out one particular Killing field. Relative to any other stationary reference system, such an experiment is in motion and potentially accelerated.
The transformation between such reference systems is similar to that known from the special theory of relativity since, at each space-time event, the four-velocities of two Killing orbits are related by a Lorentz transformation, but in general these transformations are not constant and may hence vary with space and time.

\paragraph{Time of flight and traveled distances}
Given any reference system as described above, key observables include elapsed times $T$ and traveled distances $L$ of various world-lines $\alpha \mapsto \t x{^\mu}(\alpha)$, where $\alpha$ is any quantity parameterizing the world-line. Setting $\t\Lorentz{_\mu} = - \t\Killing{_\mu}/(c^2 \lapse)$ and $\t\metric{_\mu_\nu} = \t\Metric{_\mu_\nu} + c^2 \t\Lorentz{_\mu} \t\Lorentz{_\nu}$ as above, these quantities are given by $\d T / \d \alpha = \t\Lorentz{_\mu} \d\t x{^\mu}/\d\alpha$ and $\d L / \d \alpha = \sqrt{\t\metric{_\mu_\nu} (\d\t x{^\mu}/\d\alpha) (\d\t x{^\nu}/\d\alpha)}$.
In adapted coordinates, these expressions take the form
\begin{subequations}
	\label{eq:frame dT dL}
	\begin{align}
		\label{eq:frame dT}
		\frac{\d T}{\d\alpha}
			&= \lapse \left( \frac{\d t}{\d \alpha} - \frac{\t\shift{_i}}{c} \frac{\d x{^i}}{\d \alpha} \right) \,,
		\\
		\label{eq:frame dL}
		\frac{\d L}{\d\alpha}
		   &= \sqrt{ \t\metric{_i_j} \frac{\d x{^i}}{\d\alpha} \frac{\d x{^j}}{\d\alpha}}\,,
	\end{align}
\end{subequations}
and the proper time $\tau$ along a timelike curve can be computed using $(\d \tau / \d \alpha)^2 = (\d T / \d \alpha)^2 - (\d L / \d \alpha)^2 / c^2$.

\paragraph{Space-time curvature}
\Cref{eq:frame dT,eq:frame dL} are frame-dependent as the quantities involved are derived from a specific Killing vector field $\t\Killing{^\mu}$. In situations where there are multiple Killing vector fields, it is useful to identify quantities that are frame-independent\footnote{The question of frame-independence as formulated here is vacuous for space-times admitting only a single notion of time-translations such as those discussed in Refs.~\cite{2002JHEP...10..063A,2006gr.qc....11108A,2017PhRvD..95j4039C}. Such geometries thus admit “preferred” frames and the properties of the unique (up to rescaling) Killing field are characteristic geometric invariants despite not being derived from the curvature tensor.}. Such invariant observables include  proper times of specific world-lines, but also the four-dimensional Riemann tensor $\t\Riemann{_\mu_\nu_\rho_\sigma}$ describing the curvature of the space-time metric $\t\Metric{_\mu_\nu}$. Its components can be derived from the lapse, shift, and spatial geometry. In particular, in local spatial Riemann coordinates with $\lapse = 1$ and $\t\shift{_i} = 0$ (these conditions can be satisfied simultaneously at any given point using the coordinate transformations described above), its components are given by
\begin{subequations}
\label{eq:Riem:components}
\begin{align}
	\label{eq:Riem:0i0j}
	\t\Riemann{_0_i_0_j}
		&= c^2 \t\p{_i} \t\p{_j} \lapse + \quarter \t\twist{_i_k} \t\twist{_j_l} \t\delta{^k^l}\,,
	\\
	\label{eq:Riem:0ijk}
	\begin{split}
		\t\Riemann{_0_i_j_k}
			&= \half [ (\t\p{_i} \lapse) \t\twist{_j_k} + \t\p{_i} \t\twist{_j_k} ]
			\\&\quad
			- \half [ \t\twist{_i_j} \t\p{_k} \lapse - \t\twist{_i_k} \t\p{_j} \lapse ]\,,
	\end{split}
	\\
	\label{eq:Riem:ijkl}
	\t\Riemann{_i_j_k_l}
		&= \t\riemann{_i_j_k_l} + \tfrac{3}{4} \t\twist{_i_j} \t\twist{_k_l} / c^2\,.
\end{align}
\end{subequations}
Of course, for the metrics given in \cref{eq:Minkowski:inertial,eq:Minkowski:rotation,eq:Minkowski:hyperbolic} these equations yield $\t\Riemann{_\mu_\nu_\rho_\sigma} = 0$. In particular, \cref{eq:Riem:ijkl} shows that curved spatial geometries can arise in flat space-time only relative to stationary frames of reference that have a nonzero twist tensor $\t\twist{_i_j}$.
Moreover, \cref{eq:Riem:0i0j} shows that in static reference frames (where $\t\twist{_i_j} = 0$), gravity gradients \emph{do not} directly coincide with curvature components since
\begin{align}
	\label{eq:Riem:acceleration gradient}
	\t\Riemann{_0_i_0_j} = \t\gravA{_i} \t\gravA{_j}/c^2 - \t\gamma{_i_j}\,,
\end{align}
where $\t\gravA{_i} = - \t\acc{_i}$ is the local gravitational acceleration relative to the reference system, see \cref{eq:geodesic motion x/T} below, and $\t\gamma{_i_j} = \t\p{_i} \t\gravA{_j} = \t\p{_j} \t\gravA{_i}$ is the tidal tensor.
On Earth, the gravity gradient $\t\gamma{_z_z} \approx \SI{3e-6}{\per\square\second}$ far exceeds the other term $\gravA^2/c^2 \approx \SI{e-15}{\per\square\second}$. This justifies the identification of gravity gradients with curvature in Ref.~\cite{2017PhRvL.118r3602A}. Note, however, that this identification is not universally valid: the reference system associated to \cref{eq:Minkowski:hyperbolic} has a non-zero $\t\gravA{_i}$ and $\t\gamma{_i_j}$ despite having $\t\Riemann{_0_i_0_j} = 0$.

\paragraph{Summary}
\Cref{tab:notation:space-time} summarizes the notation and interpretation of the various geometric quantities introduced so far.
Note that some of the nomenclature is not uniform throughout the literature.
In particular, the notions of lapse, shift, and spatial metric have different meanings in the \ADM\ formalism and coincide with the quantities used here only for \emph{static} reference frames \cite[Sect.~2.7]{2010_Baumgarte_Shapiro}.
Moreover, the shift (or a rescaled version) is also known as the gravitomagnetic potential \cite{2002NCimB.117..743R}, while the Lorentz-form (or variants thereof in Cartan theory) is often referred to as the clock-form \cite{2023CQGra..40j5008S}.
Some other sources, however, do not use any names for these quantities \cite[Sect.~XIV.3.2]{2009_ChoquetBruhat}.

\begin{table*}[t]
	\centering
	\begin{tblr}{@{}X[1,l]X[2.2,l]X[2.2,l]X[4.6,j]@{}}
		\toprule
		Symbol
			& Name
			& Gauge freedom
			& Significance
		\\
		\midrule
		$\t\Metric{_\mu_\nu}$
			& Space-time metric
			&
			& Determines the geometry of the space-time manifold $\Mfd$
		\\
		$\t\Killing{^\mu}$
			& Killing vector
			& $\t\Killing{^\mu} \to \text{(const.)}\, \t\Killing{^\mu}$
			& Infinitesimal generator of time-translations
		\\
		$\lapse$
			& Lapse function
			& $\lapse \to \text{(const.)} \, \lapse$
			& Measure of proper time along the orbits of $\t\Killing{^\mu}$
		\\
		$\llapse$
			& Logarithmic lapse
			& $\llapse \to \llapse + \text{const.}$
			& Analog of the Newtonian potential
		\\
		$\t\Lorentz{_\mu}$
			& Lorentz-form
			&
			& Determines the Lorentzian gamma factors of four-velocities
		\\
		$\t\acc{_i}$
			& Acceleration
			&
			& {Measure of $\t\Killing{^\mu}$-orbits deviating from space-time geodesics
			\\ Primary observable for gravimeters}
		\\
		$\t\shift{_i}$
			& Shift vector
			& $\t\shift{_i} \to \text{(const.)}(\t\shift{_i} + \t\p{_i}\lambda)$
			& Measure for the deviation from clock synchronization
		\\
		$\t\twist{_i_j}$
			& Twist tensor
			&
			& {Obstruction to clock synchronization along loops
			\\ Primary observable for Sagnac interferometers}
		\\
		$\t\metric{_i_j}$
			& Spatial metric
			&
			& Projection of $\t\Metric{_\mu_\nu}$ to the orthogonal complement of $\t\Killing{^\mu}$
		\\
		$\t\riemann{_i_j_k_l}$
			& Spatial curvature
			&
			& Measure of the deviation from Euclidean space
		\\
		$\t\Riemann{_\mu_\nu_\rho_\sigma}$
			& Space-time curvature
			&
			& {Measure of the deviation from Minkowskian space-time
			\\ Primary observable for gradiometers}
		\\\bottomrule
	\end{tblr}
	\caption{
	Summary of notation and interpretation of tensorial objects characterizing space-time and its decomposition based on a notion of time-translation invariance.}
	\label{tab:notation:space-time}
\end{table*}

\section{Phase evolution and ray trajectories}
\label{s:phase evolution}

This section shows how phase evolutions in interferometers that are “at rest” in external gravitational fields can be described concisely using the general framework of \cref{s:geometry} without the need for weak-field approximations.

\paragraph{Dispersion relations and transport laws}
The phases at two points are generally related by
\begin{align}
	\label{eq:phase evolution general}
	\Eikonal(p_2)
		&= \Eikonal(p_1)
		+ \int_\Gamma \t\WaveC{_\mu} \t{\d x}{^\mu}\,,
\end{align}
where $\Gamma$ is any curve connecting $p_1$ with $p_2$,
and $\t\WaveC{_\mu} = \t\p{_\mu} \Eikonal$ is the (angular) wave-covector.
$\t\WaveC{_\mu}$ typically satisfies a dispersion relation and a transport law along a vector field $\t\WaveV{^\mu}$ [the (angular) wave-vector, which may differ from $\t\WaveC{^\mu} \equiv \t\Metric{^\mu^\nu} \t\WaveC{_\nu}$ as is illustrated below], the explicit forms of which depend on the details of the problem.
Specifically, the analysis here is concerned with the following three cases.

\begin{enumerate}
	\item
	For freely moving massive particles, $\t\Momentum{_\mu} = \hslash \t\WaveC{_\mu}$ can be interpreted as the four-momentum. The mass $m > 0$ determines the dispersion relation $\t\Metric{^\mu^\nu} \t\Momentum{_\mu} \t\Momentum{_\nu} = - m^2 c^2$, and the corresponding transport law is $\t\Velocity{^\mu} \t\nabla{_\mu} \t\Momentum{_\nu} = 0$, where $\t\Velocity{^\mu} = \t\Metric{^\mu^\nu} \t\Momentum{_\nu} / m \equiv (\hslash/m) \t\WaveV{^\mu}$ is the four-velocity and $\t\nabla{_\mu}$ denotes the Levi-Civita derivative associated to $\t\Metric{_\mu_\nu}$. This transport equation is equivalent to the geodesic equation $\t\Velocity{^\nu} \t\nabla{_\nu} \t\Velocity{^\mu} = 0$.

	\item
	For light propagation in vacuum, the dispersion relation reads $\t\Metric{^\mu^\nu} \t\WaveC{_\mu} \t\WaveC{_\nu} = 0$. In this case $\t\WaveC{_\mu}$ is parallel-transported along the wave-vector $\t\WaveV{^\mu} = \t\Metric{^\mu^\nu} \t\WaveC{_\nu}$, i.e., $\t\WaveV{^\mu} \t\nabla{_\mu} \t\WaveC{_\nu} = 0$, which is equivalent to the geodesic equation $\t\WaveV{^\nu} \t\nabla{_\nu} \t\WaveV{^\mu} = 0$.

	\item
	For light propagation in a linear isotropic dielectric with four-velocity $\t\congr{^\mu}$ and refractive index $n$, on the other hand, $\t\WaveC{_\mu}$ satisfies the eikonal equation $\t\metric{^\mu^\nu} \t\WaveC{_\mu} \t\WaveC{_\nu} = n^2 (\t\congr{^\mu} \WaveC{_\mu})^2 / c^2$, where $\t\metric{^\mu^\nu} = \t\Metric{^\mu^\nu} + \t\congr{^\mu} \t\congr{^\nu} / c^2$ is the (contravariant) spatial metric relative to $\t\congr{^\mu}$. This is equivalent to the standard formula $k^2 = n^2 \omega^2 / c^2$ in the local rest frame of $\t\congr{^\mu}$.
	The corresponding transport law is more intricate than in the case described above: using Gordon’s optical metric $\t\MetricO{^\mu^\nu} = \t\metric{^\mu^\nu} - (n/c)^2 \t\congr{^\mu} \t\congr{^\nu}$ \cite{1923AnP...377..421G}, defined such that the dispersion relation takes the form $\t\MetricO{^\mu^\nu} \t\WaveC{_\mu} \t\WaveC{_\nu} = 0$, and denoting the corresponding Levi-Civita derivative by $\t\DelO{_\mu}$, the transport equation can be written as $\t\WaveV{^\mu} \t\DelO{_\mu} \t\WaveC{_\nu} = 0$, where $\t\WaveV{^\mu} = \t\MetricO{^\mu^\nu} \t\WaveC{_\nu}$ \cite{1967ZNatA..22.1328E}.

\end{enumerate}

Note that the equations for light propagation in vacuum can be obtained from those for massive particles using the massless limit $m \to 0$, and also from those for light propagation in media using the vacuum limit $n \to 1$.

The analysis in this paper excludes the case of guided motion of massive particles.
The reason for this omission is the lack of a comprehensive relativistic framework for modeling the phase evolution in scenarios considered in experimental proposals such as Ref.~\cite{2012PhRvL.108w0404H} or implemented in current experiments \cite{2019Sci...366..745X,2024Natur.631..515P,2025arXiv250214535D}.
While the semiclassical phase evolution of charged particles is well understood \cite{2023PhRvD.107d4029O}, such \WKB\ models currently do not apply to uncharged particles with electromagnetic dipole moments as relevant, e.g., for the description of atoms in optical lattices. On the other hand, the motion of such particles can be described using the \MPD\ equations \cite{1970RSPSA.314..499D,1970RSPSA.319..509D,1974RSPTA.277...59D} or related models \cite{2009PhRvD..80b4031G}, but these equations do not determine the phase evolution along the trajectories. Hence, further theory development is needed to provide a coherent relativistic description of such experiments.

\paragraph{Phase evolution along rays}
\Cref{eq:phase evolution general} simplifies significantly if $p_1$ and $p_2$ lie on the same flow-line of $\t\WaveV{^\mu}$.
Setting $\Gamma$ to be such a ray along which one has $\t{\d x}{^\mu} = \t\WaveV{^\mu} \d\alpha$ for some parameter $\alpha$, one obtains $\Eikonal(p_2) - \Eikonal(p_1) = \int_{\alpha_1}^{\alpha_2} \t\WaveC{_\mu} \t\WaveV{^\mu} \dd\alpha$.
Along the rays, one thus has
\begin{subequations}
\label{eq:phase evolution:light vs matter}
\begin{align}
	\label{eq:phase evolution:light}
	\d\Eikonal &= 0 &&\text{for light}\,,
	\\
	\label{eq:phase evolution:matter}
	\d\Eikonal &= - \omega_\textsc{c} \dd\tau &&\text{for matter}\,,
\end{align}
\end{subequations}
where $\tau$ denotes proper time along a timelike curve and $\omega_\textsc{c} = mc^2/\hslash$ is the (angular) Compton frequency.
It is worth noting that \cref{eq:phase evolution:light} applies both to light propagation in vacuum and to light propagation in linear isotropic media. In the latter case, the rays are \emph{timelike} and thus have $\d\tau \neq 0$, but it would be incorrect to use \cref{eq:phase evolution:matter} with $\omega_\textsc{c}$ replaced by the optical frequency relative to any observer, cf.\ Ref.~\cite[App.~A]{2017NJPh...19c3028H}.
The implications of \cref{eq:phase evolution:matter} in matter-wave gravimeters were the subject of a debate \cite{2010Natur.463..926M,2011CQGra..28n5017W,2011CQGra..28n5018S,2012Giulini,2012PhRvL.108w0404H}. In the gravimeters described in \cref{s:fountains} below, the analysis shows that the interfering rays have \emph{equal} proper times. In such cases, the Compton frequency plays no significant role in the observed phase shifts.

\paragraph{Redshift}
In general, the angular wave-covector $\t\WaveC{_\mu}$ can be decomposed as
\begin{align}
	\t\WaveC{_\mu}
		&= - \omega \t\Lorentz{_\mu}
		+ \waveC \t\waveN{_\mu}\,,
\end{align}
where $\omega$ is the angular frequency relative to the given reference system and $\t\waveN{_\mu}$ is a spatial unit covector (satisfying $\t\Killing{^\mu} \t\waveN{_\mu} = 0$ and $\t\metric{^\mu^\nu} \waveN{_\mu} \t\waveN{_\nu} = 1$). The angular wave-number $\waveC$ is then determined by the dispersion relation.
As freely-propagating light and matter are modeled as following geodesics, the quantity $\omega_* = - \t\WaveC{_\mu} \t\Killing{^\mu} = \lapse \omega$ is a constant of motion along the rays. This “angular Killing frequency” is also conserved along light rays in optical media, provided that the optical properties are time-independent.
For arbitrary points $p_i$ and $p_j$ along a ray, the redshift factor $\t z{_i_j}$, defined as $1 + \t z{_i_j} = \omega(p_j) / \omega(p_i)$, can thus be expressed in terms of the logarithmic lapse $\llapse$ as
\begin{align}
	\label{eq:redshift z}
	\begin{split}
		\t z{_i_j}
			&= \exp(\t{\Delta\llapse}{_i_j}  / c^2) - 1
			\\&
			= \t{\Delta\llapse}{_i_j}/c^2 + \half \t*{\Delta\llapse}{_i_j^2}/c^4 + O(\t*{\Delta\llapse}{_i_j^3}/c^6)
			\,,
	\end{split}
\end{align}
where $\t{\Delta\llapse}{_i_j} = \llapse(p_i) - \llapse(p_j)$.
\Cref{eq:redshift z} is an exact result in stationary space-times and the power-series expansion converges absolutely for all $\t{\Delta\llapse}{_i_j}$. In particular, this formula does not require the weak-field approximation $|\llapse| \ll c^2$.

\paragraph{Stationary solutions}

While the transport equations described above imply that $\omega_* = \lapse \omega$ is constant along the rays, this does not imply that $\omega_*$ is globally constant.
However, this additional condition is satisfied by stationary solutions, i.e., those satisfying $\Eikonal(\Phi_\chi(p)) = \Eikonal(p) - \omega_* \, \chi$ for all $p \in M$ and $\chi \in \mathbf R$ (here, $\Phi$ denotes the time-translation map introduced in \cref{s:geometry}).
This means that the frequency $\omega = \omega_*/\lapse$  relative to the orbits of $\Phi$ is constant along these timelike world-lines.
For such stationary solutions, \cref{eq:phase evolution general} reduces to a particularly simple form:
If $\Gamma$ is any curve connecting $p_1 = (t_1, x_1)$ with $p_2 = (t_2, x_2)$, one obtains
\begin{align}
	\label{eq:phase evolution stationary}
	\begin{split}
		\eikonal(x_2, x_1)
			&\coloneq
			\Eikonal(t_2, x_2) - \Eikonal(t_1, x_1) + \omega_* (t_2 - t_1)
			\\&
			= \int_\gamma k \t\waveN{_i} \t{\d x}{^i}
			+ \frac{\omega_*}{c} \int_\gamma \t\shift{_i} \t{\d x}{^i}\,,
	\end{split}
\end{align}
where $\gamma$ is the spatial projection of $\Gamma$ (a curve connecting $x_1$ with $x_2$).
The explicit form of these integrals depends on the dispersion relation:
For light one has
\begin{align}
	\label{eq:eikonal:light}
	\eikonal(x_2, x_1)
		&= \frac{1}{c} \int_\gamma n \omega\, \t\waveN{_i} \t{\d x}{^i}
		 + \frac{\omega_*}{c} \int_\gamma \t\shift{_i} \t{\d x}{^i}
		\,,
\end{align}
while for matter one has
\begin{align}
	\label{eq:eikonal:matter}
	\eikonal(x_2, x_1)
		&= \frac{1}{\hslash c} \!\int_\gamma \hspace{-0.3em} \sqrt{E^2 - m^2 c^4} \, \t\waveN{_i} \t{\d x}{^i}
		+ \frac{E_*}{\hslash c} \!\int_\gamma \hspace*{-0.3em} \t\shift{_i} \t{\d x}{^i}
		\,,
\end{align}
where $E = \hslash \omega$ is the energy (relative to the given stationary reference frame), and $E_* = \hslash \omega_* = \lapse E$ is the corresponding constant of motion.
Note that the vacuum limit $n \to 1$ in \cref{eq:eikonal:light} produces the same result as the massless limit $m \to 0$ in \cref{eq:eikonal:matter}.

\paragraph{Ray trajectories}
Full integration of \cref{eq:phase evolution:light,eq:phase evolution:matter} or \cref{eq:eikonal:light,eq:eikonal:matter} requires knowledge of the rays.
Here, we distinguish two cases: (i) guided motion and (ii) free motion.
Guided motion (i) applies, for example, to light propagating through optical fibers. In this case, the spatial trajectories are determined by the experimental setup and the temporal part of the corresponding rays is determined by the local speed of light, i.e., by the fiber’s refractive index.
Free motion (ii) describes freely falling matter and also free light propagation in vacuum, in which case the rays are modeled as space-time geodesics. Their spatial projections satisfy the equation
\begin{multline}
	\label{eq:geodesic motion x/T}
	\frac{\d^2 \t x{^i}}{\d T^2}
		+ \Christoffel{i}{j}{k} \frac{\d\t x{^j}}{\d T} \frac{\d\t x{^k}}{\d T}
		\\=
		\left(
			\t\metric{^i^j}
			- \frac{1}{c^2} \frac{\d\t x{^i}}{\d T} \frac{\d\t x{^j}}{\d T}
		\right) \t\gravA{_j}
		- \t\twist{^i_j} \frac{\d\t x{^j}}{\d T}\,,
\end{multline}
where $T$ is a time-parameter along the curve that is defined by \cref{eq:frame dT dL}, i.e., $\d T = \t\Lorentz{_\mu} \d\t x{^\mu}$. Here, $\christoffel{i}{j}{k}$ denote the Christoffel symbols of the spatial metric $\t\metric{_i_j}$ and $\t\gravA{_i} = - \t\acc{_i} = - \t\p{_i} \llapse$ is the negative acceleration of the stationary Killing orbits.
This equation shows that massive particles that are initially at rest in the considered reference frame experience an acceleration that is equal to $\t\gravA{_i}$. Note that, despite this close analogy to Newtonian gravity, \cref{eq:geodesic motion x/T} is an exact result that holds in arbitrary stationary space-times.
As such, this equation generally does not admit closed-form solutions, implying that exact or approximate solutions for specific experiments must be computed on a case-by-case basis.
Explicit expressions are discussed in \cref{s:experiments:light,s:experiments:matter}.

\section{Detection probabilities from phase shifts}
\label{s:quantum probabilities}

The methods outlined in \cref{s:phase evolution} describe the deterministic evolution of the phase function $\Eikonal$.
While the phase is generally considered as being not directly observable, \emph{phase differences} arising from interfering rays determine detection probabilities. This section summarizes the main equations relevant to the description of interferometric gravimeters and gradiometers using single-particle and multi-particle states following either bosonic or fermionic statistics.

For a general interferometer with $N_\text{in}$ inputs and $N_\text{out}$ outputs, let $\ket{\text{in}_i}$ denote a (normalized) single-particle excitation of the input mode $i$. In the Heisenberg picture, this state \emph{equals} a superposition of single-particle excitations $\ket{\text{out}_{k|i}}$ of the various output modes $k$. Modeling potential losses using an additional “fictitious output port” that remains undetected, one thus has
\begin{align}
	\ket{\text{in}_i}
		= \sum_{k = 1}^{N_\text{out}} \ket{\text{out}_{k|i}} + \ket{\text{loss}_{i}}\,.
\end{align}
The corresponding equation for ladder operators takes the form%
\footnote{%
	\Cref{eq:ladder operator decomposition} is reminiscent of Eq.~(8) in Ref.~\cite{2023AnP...53500468B} and Eq.~(39) in Ref.~\cite{2023JPhCS2531a2016A}, both of which relate ladder operators corresponding to different frequency spectra rather than spatial modes.
	The criticism of these equations expressed in Ref.~\cite{2025arXiv250220521L} does not apply to \cref{eq:ladder operator decomposition} as this equation describes a transformation of the structure $N \to N^2 + N$ [expressing the $N$ input-operators in terms of $N^2$ output-operators and $N$ loss-operators], rather than $N \to N$ as considered in Refs.~\cite{2023AnP...53500468B,2023JPhCS2531a2016A}.%
}%
\begin{align}
	\label{eq:ladder operator decomposition}
	\t* a{^\dagger_i} = \sum_{k} \t*b{^\dagger_k_|_i} + \t*c{^\dagger_i}\,.
\end{align}
By construction, the input vectors $\ket{\text{in}_i}$ are orthonormal, and the output vectors satisfy $\braket{\text{out}_{k|i} \mid \text{out}_{l|j}} = 0$ for $k \neq l$ since excitations at different outputs are perfectly distinguishable. For $k = l$, however, the inner products
\begin{align}
	\t u{_k_|_i_j}
		&= \braket{\text{out}_{k|i} \mid \text{out}_{k|j}}
		\equiv \braket{0|\t*b{_k_|_i} \t*b{^\dagger_k_|_j} |0}
\end{align}
have a non-trivial structure that depends on the details of the interferometer.
A simple model of a Mach–Zehnder interferometer with symmetric $50:50$ beam splitters and equal losses in both arms is obtained by setting
\begin{subequations}
\begin{align}
	(\t u{_1_|_i_j})
		&=
		\half \mathscr A
		\begin{pmatrix}
				1 - \mathscr V \cos\Delta\Eikonal
			&	\phantom1 + \mathscr V \sin\Delta\Eikonal
			\\
				\phantom1 + \mathscr V \sin\Delta\Eikonal
			&	1 + \mathscr V \cos\Delta\Eikonal
		\end{pmatrix}\,,
	\\
	(\t u{_2_|_i_j})
		&=
		\half \mathscr A
		\begin{pmatrix}
				1 + \mathscr V \cos\Delta\Eikonal
			&	\phantom1 - \mathscr V \sin\Delta\Eikonal
			\\
				\phantom1 - \mathscr V \sin\Delta\Eikonal
			&	1 + \mathscr V \cos\Delta\Eikonal
		\end{pmatrix}\,,
\end{align}
\end{subequations}
where $\mathscr A$ and $\mathscr V$ are parameters describing the overall transmission probability and the fringe visibility, respectively, and $\Delta\Eikonal$ takes the form
\begin{align}
	\Delta\Eikonal
		= \Delta\Eikonal_\lapse + \Delta\Eikonal_\shift + \Delta\Eikonal_*\,,
\end{align}
where $\Delta \Eikonal_\lapse$ and $\Delta \Eikonal_\shift$ depend, respectively, only on the lapse and the shift along the rays $\gamma_\I$ and $\gamma_\II$ as
\begin{subequations}
	\label{eq:phase shifts lapse and shift}
	\begin{align}
		\Delta \Eikonal_\lapse &
			= \int_{\gamma_\I} \t\waveC{_\I} \t\waveN{_i} \t{\dd x}{^i}
			- \int_{\gamma_\II} \t\waveC{_\II} \t\waveN{_i} \t{\dd x}{^i}\,,
		\\
		\Delta \Eikonal_\shift &
			= \frac{\omega_*}{c} \int_{\gamma_*} \t\shift{_i} \t{\d x}{^i}\,.
	\end{align}
\end{subequations}
Here, the wave vectors $\t\waveC{_\I}$ and $\t\waveC{_\II}$ depend on the dispersion relation, see \cref{eq:eikonal:light,eq:eikonal:matter}, and $\gamma_*$ denotes the closed path obtained by concatenating $\gamma_\I$ with the reverse of $\gamma_\II$.
[If $\gamma_*$ encloses an area $A$, Stokes’ theorem implies $\Delta\Eikonal_\shift = \int_A \omega \, \t\twist{_i_j} \t{\d x}{^i} \t{\d x}{^j} / c^2$.]
Finally, $\Delta\Eikonal_*$ accounts for all additional phase shifts arising, for example, from optical mirrors and phase shifters (which are commonly used to introduce adjustable bias phases) or from the interaction of massive particles with matter or electromagnetic radiation \cite{2001Metro..38...25P}. Explicit expressions for $\Delta\Eikonal$ depend on the experiment under consideration, see \cref{s:experiments:light,s:experiments:matter} below.

For both  bosonic and fermionic particles, the probability of detecting one particle at output no.~1 after it being injected into input no.~1 is given by
\begin{align}
	\mathrm P_1
		&= \half \mathscr A [1 - \mathscr V \cos(\Delta\Eikonal)]\,.
\end{align}
The coincide probability of detecting one particle at each of the outputs after two identical particles have been injected at the two inputs depends on the statistics:
\begin{subequations}
\begin{align}
	\mathrm P_2^\text{bos.}
		&= \half \mathscr A^2 [1 + \mathscr V^2 \cos(2\Delta\Eikonal)]\,,
	\\
	\mathrm P_2^\text{ferm.}
		&= \half \mathscr A^2 [1 + \mathscr V^2]\,.
\end{align}
\end{subequations}
In general, gravity influences both the phase shift $\Delta\Eikonal$ and the visibility $\mathscr V$.
However, since the former is typically more sensitive to gravitational effects than the latter \cite{2012CQGra..29v4010Z,2022PhRvA.106c1701M,2023Mieling}, the following analysis focuses on $\Delta\Eikonal$ and treats $\mathscr V$ as a constant.

\Cref{eq:phase shifts lapse and shift} shows that, in the current model, all detection probabilities are expressible in terms of integrals of the lapse and shift along rays. Since other geometric quantities such as the curvature tensor $\t\Riemann{_\mu_\nu_\rho_\sigma}$ do not enter explicitly, information about them can only be obtained by indirect means, e.g., using \cref{eq:Riem:components}.
This absence of a direct curvature coupling is the result of the \WKB\ approximation underlying the calculations of \cref{s:phase evolution}: more detailed analyses in Refs.~\cite{2021CmPhy...4..171E,2020PhRvD.102b4075O,2025PhRvR...7a3162M} show that Maxwell’s equations imply curvature-induced corrections to the dispersion relation, but such higher-order corrections are inaccessible to current experiments and are thus beyond the scope of this paper.

Having laid out the general methods for computing detection probabilities for quantum interferometers in a gravitational field, \cref{s:experiments:light,s:experiments:matter} describe concrete experiments involving massless and massive particles, respectively.

\section{Optical experiments}
\label{s:experiments:light}

This section does not follow the chronological development of the subject but rather present various experimental proposals sorted according to the mathematical complexity of their models.
For simplicity, the calculations presented here mainly focus on \emph{static} frames of reference and treat shift-induced effect perturbatively.

\subsection{Pound–Rebka experiment}
\label{s:Pound Rebka}

In their 1960 experiment, Pound and Rebka determined the gravitational redshift in γ-rays (that were emitted in the decay of \textsuperscript{57}Co to \textsuperscript{57}Fe) by measuring the difference in frequency over a height difference of $h = \SI{74}{\foot} \approx \SI{22.6}{\meter}$ \cite{1960PhRvL...4..337P}. The result was found to be compatible with the theoretical prediction
\begin{align}
\label{eq:redshift}
	\begin{split}
		z
			&= \frac{\nu_h - \nu_0}{\nu_0}
			\approx \Delta \llapse_{0\,h} / c^2
			\\&
			\approx - \gravA h /c^2
			\approx \num{- 2.46e-15}\,,
	\end{split}
\end{align}
which is a direct application of \cref{eq:redshift z}.
Since this experiment measured the shift in γ-ray frequencies relative to nuclear transition frequencies \cite{2000AmJPh..68..115O}, a debate arose whether the redshift is to be interpreted as a gravitational effect on light—as suggested by the paper title “Apparent Weight of Photons” \cite{1960PhRvL...4..337P}—or a gravitational effect on the emitting and absorbing atoms—as suggested by the title of the follow-up paper “Effect of Gravity on Nuclear Resonance” by Pound and Snider \cite{1964PhRvL..13..539P}—see, e.g., Ref.~\cite{2000AmJPh..68..115O} and references therein.
Note, however, that the current definition of the second \cite{CGPM26Resolution1} fixes the frequency of light emitted in the ground-state hyperfine transition of {}\textsuperscript{133}Cs. Consequently, in a (hypothetical) variant of the Pound–Rebka experiment involving such radiation, gravitational effects on the emission frequencies of the atoms are excluded \emph{by definition}, thus leading to the alternative description of the redshift as a gravitational effect on light propagation.

The prevalent view that the Pound–Rebka experiment can be described—on the whole or at least in its central aspects—classically, i.e., without quantum theory \cite{2012CQGra..29v4010Z,2015PhRvD..91f4041B,2016JPhCS.723a2044Z,2017NJPh...19e0401A,2024PhRvL.133b0201W,2017EPJQT...4....2P}, has motivated the development of alternative experiments that could distinguish more clearly between the predictions of quantum optics in curved space-times and its classical or flat limits.

\subsection{Tanaka’s proposal}
\label{s:Tanaka}
\begin{figure}[t]
	\centering
	\includegraphics[page=1]{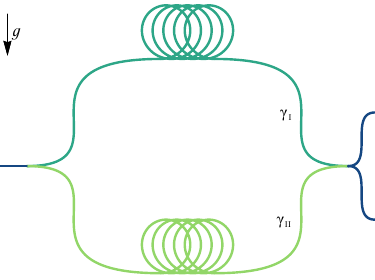}
	\caption{
		Simplified schematic of the fiber-optic interferometer proposed in Ref.~\cite{1983PhRvL..51..378T} where a light ray is split into two paths that traverse fiber spools of equal length. The resulting phase shift at the output depends on the vertical separation between the two fiber spools via the redshift.}
	\label{fig:tanaka}
\end{figure}

In his 1983 paper, Tanaka suggested using a fiber-optic Mach–Zehnder interferometer for measuring a phase shift that is induced on light by the gravitational redshift \cite{1983PhRvL..51..378T}. A simplified schematic representation of such an interferometer is provided in \cref{fig:tanaka}.
In this setup, a coherent beam of light is injected on the left and split into two beams that traverse fiber-optic spools at different heights (the two spatial trajectories are denoted by $\gamma_\I$ and $\gamma_\II$, respectively), after which the rays interfere at a fiber-coupler before being detected. Rotating the interferometer about the horizontal allows modulating the gravitational phase difference between the two rays following the prescribed trajectories.
Based on \cref{eq:phase evolution stationary,eq:eikonal:light}, one finds the lapse-induced phase shift for stationary solutions to be given by
\begin{align}
	\Delta \Eikonal_\lapse
		&= \frac{n \omega_*}{c} \left[
			\int_{\gamma_\I} \frac{\d \length}{\lapse}
			- \int_{\gamma_\II} \frac{\d \length}{\lapse}
		\right]\,,
\end{align}
where $\omega_*$ is the Killing frequency.
This ray-optics result agrees with more rigorous fiber-optics calculations up to higher-order curvature corrections that are negligible in most applications \cite{2018CQGra..35x4001B,2022PhRvA.106f3511M,2025PhRvR...7a3162M}.
If both fiber spools have equal length $\length$ and the lapse is approximately constant along each of the two spools, the phase shift $\Delta \Eikonal_\textsc{t}$ arising from small vertical separations $h$ can be approximated by
\begin{align}
	\label{eq:phase:Tanaka}
	\begin{split}
		\Delta \Eikonal_\textsc{t}
			&\approx - \gravA h \length n \omega / c^3
			\\&
			\approx - \SI{e-6}{\radian}
			\, (h / \si{\meter})
			\, (\length / \si{\kilo\meter})
			\, (\lambda_0 / \si{\micro\meter})^{-1}\,,
	\end{split}
\end{align}
where $n \approx 1.5$ is the refractive index of glass and $\lambda_0 = 2 \pi c / \omega$ is the wavelength of light in vacuum, see also Ref.~\cite{2017NJPh...19c3028H}.

After Ref.~\cite{1983PhRvL..51..378T} originally suggested performing the experiment with continuous-wave laser output, more recent papers suggested using single photons, pairs of entangled photons, and other non-classical states of light instead \cite{2017NJPh...19c3028H,2019PhRvA..99b3803C,2022PhRvA.106f3511M,2022PhRvA.106c1701M}.
With appropriately chosen bias phases [specifically $\pi/(2 N)$ for $N$-photon N00N-states], and if the Sagnac effect is negligible, the single-photon and two-photon detection probabilities take the form
\begin{subequations}
	\begin{align}
		\mathrm P_1
			&\approx \half \mathscr A [1 - \mathscr V \gravA h \length n \omega/c^3 ]\,,
		\\
		\mathrm P_2
			&\approx \half \mathscr A [1 + 2 \mathscr V^2 \gravA h \length n \omega/c^3 ]\,.
	\end{align}
\end{subequations}
Whereas the single-photon detection probabilities are identical to classical field intensities \cite{2022PhyS...97k4004B}, the non-classical factor of two in $\mathrm P_2$ makes the two-photon coincidence experiment immune to such criticism.
As a consequence, two-photon interferometry experiments of this kind could demonstrate gravitationally induced quantum interference without a classical analog.

In Ref.~\cite{1983PhRvL..51..378T}, such an experiment was interpreted as a “twin paradox for photons.”
However, contrary to the standard “twin paradox,” the rays interfering at the second fiber coupler (where the world-lines intersect) \emph{do not} pass the first fiber coupler at the same time. Indeed, the phase shift given above is fully attributable to such a temporal offset \cite{2012CQGra..29v4010Z,2023Mieling}. This is consistent with \cref{eq:phase evolution:light}, which shows that the rays acquire no gravitational phases while propagating through the fibers.

While the formulas given here directly apply to static frames of references, their use in stationary frames (where $\t\shift{_i} \neq 0$) requires some caution because the shift-induced phase difference $\Delta\Eikonal_\shift$ is generally non-negligible. Using \cref{eq:phase shifts lapse and shift} the ratio $|\Delta\Eikonal_\shift / \Delta\Eikonal_\lapse|$ can be estimated by $2 (\cos \vartheta) c A \Omega / (\gravA \length h)$, where $A$ is the effective area of the surface enclosed by the interferometer and $\vartheta$ is the angle between the corresponding surface normal and the local rotation vector $\t\Omega{^i}$.
This shows that, on Earth, the Sagnac phase shift exceeds the gravitationally induced one by a factor of about $\num{4e3} (\cos \vartheta) A/(\length h)$. As a consequence, the isolation of $\Delta\Eikonal_\lapse$ poses a severe experimental challenge. (For recent experiments on the Sagnac effect in quantum fiber-interferometry, see, e.g., Refs.~\cite{2006JPhB...39.1011B,2022PhRvL.129z0401T,2023PhRvR...5b2005C,2024SciA...10O.215S}. Their detailed description, however, is beyond the scope of this paper).

\subsection{Stodolsky’s proposal}
\label{s:Stodolsky}
\begin{figure}[t]
	\centering

	\subfloat[]{\includegraphics[page=2]{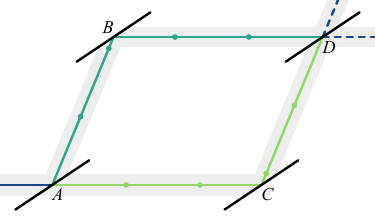}}

	\subfloat[]{\includegraphics[page=3]{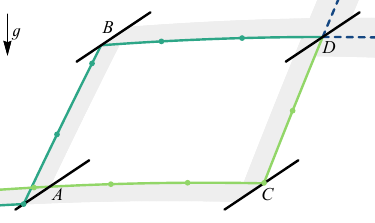}}

	\caption{
		Schematic representation of optical free-fall trajectories in a Mach–Zehnder interferometer, with dots marking which points along the two rays reach the point \textit{D} simultaneously.
		If the system is inertial, interfering rays at \textit{D} originate from the same point on the beam splitter \textit{A}, which they pass at the same time  (a).
		This no longer applies in uniformly accelerated interferometers, as is the case in gravimeters (b).
		Instead, the interfering rays must have a spatial and temporal offset at the first beam splitter \textit{A} in order to reach the final beam splitter at the same space-time event.
		The gray area indicates the minimal beam width required to observe the gravitationally induced phase shift.
	}
	\label{fig:MZI setups}
\end{figure}

In his 1979 paper, Stodolsky proposed a free-space Mach–Zehnder interferometer for measuring the gravitationally induced phase shift of light \cite{1979GReGr..11..391S} (this was presented as an optical variant of the experiment done by Colella, Overhauser, and Werner, which is described in \cref{s:COW} below, and predates Tanaka’s proposal described above).
A simplified schematic of the setup is provided in \cref{fig:MZI setups}. In contrast to Tanaka’s setup, light is not guided through optical fibers, but follows free-fall trajectories between the beam splitters and the mirrors.
Explicit integration of the geodesic equation \eqref{eq:geodesic motion x/T} in spatial Riemann coordinates with constant $\t\gravA{_i}$ shows that the rays interfering at the second beam splitter (\textit{D}) pass the first beam splitter (\textit{A}) at different times and different positions. To leading order in the acceleration $\gravA$, these offsets are given by
\begin{subequations}
	\label{eq:offsets:Stodolsky}
	\begin{align}
		\Delta t_\textsc{s} &= - (\gravA h \length/c^3)(\cot^2\theta - 1)\,,
		\\
		\Delta x_\textsc{s} &= - (\gravA h \length/c^2) \cot^2 \theta\,,
		\\
		\Delta z_\textsc{s} &= - (\gravA h \length/c^2) \cot \theta\,,
	\end{align}
\end{subequations}
where $\theta$ is the inclination of the mirrors and beam splitters that are separated horizontally by the length $\length$ and vertically by the height $h$.
Here and in the following, $\t{\Delta x}{^\mu} = \t*x{^\mu_\I} - \t*x{^\mu_\II}$ refers to the coordinate separation between the upper ray $\t\gamma{_\I}$ (\textit{ABD}) and the lower ray $\t\gamma{_\II}$ (\textit{ACD}) at the respective events of entering the interferometer\footnote{The coordinate displacement arising here can be formalized geometrically using the notion of geodesic separation vectors, but this is not necessary for the purpose of computing the difference in phase between the two events with coordinates $\t*x{^\mu_\I}$ and $\t*x{^\mu_\II}$ using \cref{eq:phase evolution general}.}.
Since the phases are constant along the rays, the phase shift at \textit{D} is given by the “separation phase” $\Eikonal(\t*x{_\I}) - \Eikonal(\t*x{_\II}) \approx \t\WaveC{_\mu}\t{\Delta x}{^\mu}$, see \cref{eq:phase evolution general}. Using $(\t\WaveC{_\mu}) \approx (- \omega, \omega/c, 0, 0)$, the lapse-induced phase shift thus takes the form
\begin{align}
	\label{eq:phase:Stodolsky}
	\Delta \Eikonal_\textsc{s}
		&\approx - \gravA h \length \omega / c^3\,,
\end{align}
which is identical to the phase shift in Tanaka’s setup with $n = 1$. Hence, even though the two setups differ in the details of light propagation (guided motion within optical fibers is generally not geodesic), the resulting phase shifts and detection probabilities coincide.

The point of view that \emph{geodesic motion} of light in the interferometer plays no essential role is supported by comparison with the analysis of Ref.~\cite{2012CQGra..29v4010Z}.
The rays considered there are lightlike curves of constant radial coordinate $r$ in the Schwarzschild chart. Such rays fail to be geodesic unless $r = 3 G M /c^2$ \cite[Sect.~6.3]{1984_Wald}, a condition that rules out the comparison of two such rays at different $r$ and is impossible to satisfy on Earth (the photon sphere of the Schwarzschild geometry with $M$ equal to Earth’s mass would lie at $r \approx \SI{1.3}{\centi\meter}$).
Nonetheless, the resulting phase shift agrees with \cref{eq:offsets:Stodolsky} to the considered level of approximation.
However, these observations do not imply that the geometric details of the interferometer are wholly irrelevant.
Indeed, as shown in Appendix~\ref{s:linearized gravity Stodolsky}, imbalances in the arm lengths can significantly alter the phase shift. For example, the assumption that the interferometer arms have equal \emph{coordinate lengths} in the frequently used post-Newtonian gauge corresponds to having unequal \emph{physical lengths} and hence leads to deviations from \cref{eq:phase:Stodolsky}, see also Ref.~\cite{2015PhRvD..91f4041B}.

Similarly to Tanaka’s setup, the relative size of the shift-induced phase difference can be estimated by the ratio $2 (\cos \vartheta) c \Omega/g \approx \num{4e3} \cos \vartheta$. The isolation of the lapse-induced phase shift thus requires special alignments relative to the local rotation axis or alternative compensation or subtraction schemes.

\begin{figure*}
	\centering

	\subfloat[]{%
		\includegraphics[page=4,width=0.6\columnwidth]{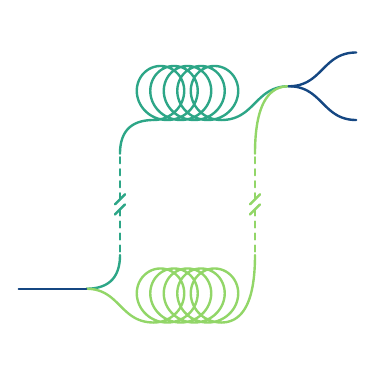}%
		\label{fig:satellite single Mach-Zehnder}%
	}
	\hfill
	\subfloat[]{%
		\includegraphics[page=5,width=0.6\columnwidth]{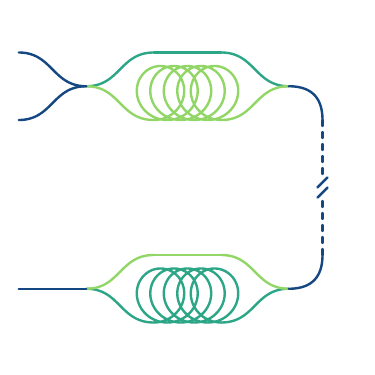}%
		\label{fig:satellite double Mach-Zehnder}%
	}
	\hfill
	\subfloat[]{%
		\includegraphics[page=6,width=0.6\columnwidth]{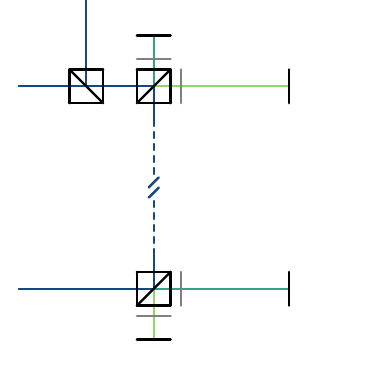}%
		\label{fig:satellite double Michelson}%
	}

	\caption{
		Schematic representation of satellite experiments to measure the gravitationally induced phase shift on light.
		The setup \ref{fig:satellite single Mach-Zehnder} is a large-scale version of Tanaka’s setup (\cref{s:Tanaka}) and requires two uplinks.
		The setups \ref{fig:satellite double Mach-Zehnder} and \ref{fig:satellite double Michelson}, on the other hand, use a scheme similar to time-bin-encoding for which a single uplink suffices.
		All these setups, however, yield the same lapse-induced phase shift.
	}
	\label{fig:satellite setups}
\end{figure*}

\subsection{Proposed Satellite Experiments}
In 2012, Rideout \etal.\ proposed measuring the gravitationally induced phase shift on light in a satellite experiment \cite{2012CQGra..29v4011R}. Subsequent papers described variants of this proposal \cite{2022PhRvA.106c1701M,2024PhRvL.133b0201W,2020PhRvD.101j4052T,2022EPJQT...9...25M,2022PhRvA.106c1701M,2022PhRvA.106f3511M,2023PhRvD.108h4063T,2025RvMP...97a5003B,2022RvMP...94c5001L,2024PhRvL.133b0201W}, whose setups are sketched in \cref{fig:satellite setups}.
In all considered cases, the lapse-induced phase shift takes the form
\begin{align}
	\begin{split}
		\Delta \Eikonal_\lapse
			&= \frac{n \omega_* \length}{c}
			\left[ \frac{1}{\lapse_\text{s}} - \frac{1}{\lapse_\text{g}} \right]
			\\&
			= n \length \omega_\text{g} z_{\text{g} \text{s}} / c
			\approx n \length \omega_\text{g} (\llapse_\text{g} - \llapse_\text{s}) / c^3\,,
	\end{split}
\end{align}
where $n$ is the refractive index of the fiber-optic delay lines in the setups \ref{fig:satellite single Mach-Zehnder} and \ref{fig:satellite double Mach-Zehnder}, and $n = 1$ in the setup \ref{fig:satellite double Michelson}.
The parameter $\length$ denotes the length of the delay lines, $\omega_\text{g}$ denotes the optical frequency at the ground station, and $z_{\text{g} \text{s}}$ is the redshift factor \cite{2012CQGra..29v4011R,2022EPJQT...9...25M,2022PhRvA.106c1701M,2024PhRvL.133b0201W,2025RvMP...97a5003B}.
However, the lapse-induced phase shift is not the only relevant phase contribution in satellite experiments \cite{2016PhRvL.116y3601V}.
Indeed, in setups of the type \ref{fig:satellite single Mach-Zehnder} the large area enclosed by the interferometer produces a significant shift-induced phase difference $\Delta \Eikonal_\shift$, and, moreover, in all setups shown in \cref{fig:satellite setups} the satellite’s motion gives rise to an additional phase shift $\Delta\Eikonal_\textsc{d}$ via the Doppler effect \cite{2020PhRvD.101j4052T}.
Future implementations of such experimental setups thus require compensation schemes to isolate $\Delta \Eikonal_\lapse$ \cite{2023PhRvD.108h4063T,2024PhRvL.133b0201W}.

\subsection{Main Experimental Challenges}

All optical experiments described above ultimately depend on the magnitude of the redshift as given by \cref{eq:redshift z}. A direct observation of this effect with single photons requires either a large height difference between the source and receiver, or the use of short wavelengths, as was done in the Pound–Rebka experiment. For example, at telecommunication wavelengths (\SI{1550}{\nano\meter}) the absolute value of the gravitational frequency shift is about \num{18000} times smaller than that for the γ-rays used in the Pound–Rebka experiment (\SI{86}{\pico\meter}). In interferometric experiments using single- or multi-photon states, the gravitational redshift manifests itself as a phase shift between the paths at different altitudes. By extending the path lengths, for example using optical fibers, this phase shift becomes detectable even at telecom wavelengths and laboratory-scale height differences \cite{2012CQGra..29v4010Z,2017NJPh...19c3028H}.

A key experimental challenge is the isolation of the gravitational signal from phase noise introduced by the light-guiding medium. Any differential change in the optical path length leads to phase shifts that cannot be distinguished from the signal of interest within the same frequency band. To stabilize the interferometer arms, a coherent light source at a frequency distinct from that of the photons in the quantum state can be used. This wavelength-division multiplexing method allows for two different approaches:

\begin{enumerate}
	\item The stabilization laser can be used to maintain a constant phase difference between the two arms, regardless of their vertical separation. This approach compensates for the gravitational redshift experienced by the laser and thus only allows to bound any potential difference in the gravitationally induced phase shifts between the laser and the single- or multi-photon quantum states, i.e., the quantity $\epsilon = \Delta \Eikonal_\lapse^\text{laser} - \Delta \Eikonal_\lapse^\text{photon}$. (Note, however, that the theoretical prediction $\epsilon = 0$ is independent of the value of $\gravA$ and, moreover, does not depend on the validity of Einstein’s field equations.)
	\item Alternatively, after initial calibration, the two interferometer arms can be stabilized independently. This enables direct measurements of the lapse-induced phase shift of the quantum states as a function of height, assuming sufficient  phase stability in both arms independently.
\end{enumerate}

Independent stabilization places severe bounds on laser stability and accuracy. This can be exemplified by the satellite experiments using time-bin encoding shown in \cref{fig:satellite double Mach-Zehnder,fig:satellite double Michelson}. The first constraint arises because different lasers are used to stabilize the two unbalanced interferometers, one at the ground station and one on the satellite. The difference in their central frequencies, $\Delta \nu$, must not exceed the gravitational redshift on the signal photons, $|\Delta \nu| < \nu_0 |\Delta\llapse|/c^2$. For experiments on Earth, this is a severe constraint as it leads to an admissible difference in laser frequencies of only a few tens of \si{\milli\hertz} per meter of height difference while for satellite experiments the acceptable difference relaxes to several \si{\kilo\hertz}. The second constraint arises from the wavelength differences between the stabilization lasers and the photons whose redshift is to be measured. In each unbalanced Mach–Zehnder interferometer, the induced phase difference between the stabilization laser and the quantum-state photons varies with the arm length difference $\delta l$ between the long arm common to both rays and the short reference arm. Any difference in length imbalance between the ground ($\delta l_\text{g}$) and the satellite station ($\delta l_\text{s}$) leads to a phase shift between the interfering time-bin amplitudes and must be smaller than the gravitational phase, $2 \pi n |\nu_0-\nu_\text{laser}|\,|\delta l_\text{s}-\delta l_\text{s}|/c < |\Delta \Eikonal_\lapse|$. This constraint is much less demanding than the one mentioned above, but still requires attention for earthbound experiments. For \SI{10}{\kilo\meter}-long fiber spools separated in altitude by \SI{20}{\meter}, the frequency difference between telecom-wavelength photons and the laser should be smaller than about \SI{4}{\mega\hertz} for $|\delta l_\text{g}-\delta l_\text{s}| = \SI{1}{\milli\meter}$. Since this is too narrow for filtering, the stabilization laser needs to be injected from the other end of the interferometer.

Although optical fibers allow for a comparatively simple way of increasing the interferometric area by extending the optical path length in each arm, they also introduce significant challenges. Losses in the fibers and other optical components can severely limit the number of detected photons and thus the precision in the estimation of the phase shift due to shot noise. High losses thus increase the duration of the measurement and therefore the stability requirements on the experiments. The simultaneous use of quantum states and stabilization lasers within the same optical fiber can also introduce noise via several mechanisms \cite{2022Optic...9.1238H}. Spontaneous Raman scattering of photons from the laser into the quantum channel and fundamental thermal noise are among the most difficult to deal with. A potential solution is to use (doubly) nested anti-resonant nodeless fibers (\textsc{dnanf}) \cite{2014OExpr..2223807P} instead of standard single mode fibers. In \textsc{dnanf}, more than \SI{99}{\percent} of the optical mode can propagate in vacuum with greatly reduced losses \cite{Chen:24} and non-linear effects \cite{2019JLwT...37..909L}.

Several methods exist to separate the lapse-induced phase from phase shifts induced by the Sagnac effect. The conceptually simplest approach is to orient the interferometer such that its surface normal is aligned at an angle of $\vartheta=\pi/2$ with respect to Earth’s rotation axis. While straightforward in principle, this configuration is only practical for small-scale, laboratory-based experiments. For schemes using laser stabilization to maintain a constant phase relationship between the two arms, the Sagnac effect is common to both the coherent state and the single- or multi-photon quantum states and thus does not contribute to the differential output signal. When optical fibers are used, they can be coiled so that the winding direction is reversed halfway through the fiber length. This configuration ensures that light in both halves of the fiber experiences equal and opposite phase shifts, resulting in a net-zero Sagnac contribution.

\section{Matter-wave experiments}
\label{s:experiments:matter}

Whereas gravitationally induced phase shifts have not yet been observed in optical experiments, analogous effects in matter-wave experiments have already been measured. This section describes the main characteristics of these experiments using the general methods developed in \cref{s:phase evolution,s:quantum probabilities}.
Similarly to the analyses in \cref{s:experiments:light} above, the calculations presented here mainly focus on static frames of references for simplicity of exposition.

\subsection{Colella–Overhauser–Werner experiment}
\label{s:COW}
\begin{figure}[t]
	\centering

	\subfloat[]{%
		\includegraphics[page=7]{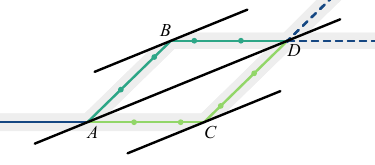}
		\label{fig:OC-gravity}
	}

	\subfloat[]{%
		\includegraphics[page=8]{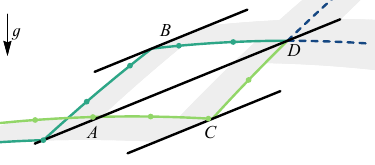}
		\label{fig:OC+gravity}
	}

	\caption{
		Schematic representation of the neutron interferometer proposed by Overhauser and Colella in Ref.~\cite{1974PhRvL..33.1237O}.
		In the absence of gravity the rays form a rhombus (a), but in the presence of gravity rays interfering at $D$ must have been injected into the interferometer at different heights (b).
		The dots indicate which points along the two rays reach the point \textit{D} simultaneously and the gray region shows the minimal beam width required to observe phase shifts in the presence of gravity.
	}
	\label{fig:OC}
\end{figure}

In their 1974 paper, Overhauser and Colella proposed a neutron-interferometry experiment for demonstrating gravitationally induced phase shifts \cite{1974PhRvL..33.1237O}, a simplified schematic of which is shown in \cref{fig:OC}.
In this setup, a neutron beam is split, reflected, and recombined by diffraction at three parallel slabs cut from a single crystal of silicon.
The expected phase shift in the Overhauser–Colella setup is given by
\begin{align}
	\label{eq:OC:phase shift}
	\begin{split}
		\Delta \Eikonal_\textsc{oc}
			&= - \frac{\gravA h l}{\hslash \velocity^2} \frac{m \velocity}{\sqrt{1 - \velocity^2/c^2}}
			\\&
			= - \frac{\gravA h l m}{\hslash \velocity} \left[1 + O(\velocity^2/c^2)\right]\,,
	\end{split}
\end{align}
where the height difference $h$ depends on the base length $\length$ and the inclination angle $\theta$ of the crystal slabs, $h = \length \sin(2 \theta)$, and $\velocity$ is the neutron’s velocity.
The leading-order term in \cref{eq:OC:phase shift} was originally derived using the semiclassical approximation of the Schrödinger equation and assuming the neutron rays to be straight lines \cite{1974PhRvL..33.1237O}. Similar derivations of the phase shift using straight rays are given, for example, in Refs.~\cite{1977PhRvD..15.1448A,1993PhLA..182..330K,2000AmJPh..68..404V,2017GReGr..49...82G,2025RvMP...97a5003B}.
In his 1998 paper, Mannheim showed that—by taking into account the bending of the rays under the influence of gravity—the phase shift can be explained by the fact that rays interfering at the output must have reached the central crystal slab at different times \cite{1998PhRvA..57.1260M}. Specifically, the temporal and spatial offsets of the rays shown in \cref{fig:OC+gravity} are given by
\begin{subequations}
	\label{eq:OC:offsets}
	\begin{align}
		\Delta t_\textsc{oc}
			&= - \frac{\gravA h l}{\velocity^3} \left[\cot^2 \theta + 1 - 2 \frac{\velocity^2}{c^2}\right]\,,
		\\
		\Delta x_\textsc{oc}
			&= - \frac{\gravA h l}{\velocity^2} \cot^2 \theta\,,
		\\
		\Delta z_\textsc{oc}
			&= - \frac{\gravA h l}{\velocity^2} \cot \theta\,,
	\end{align}
\end{subequations}
as can be calculated by explicitly integrating the geodesic equation \eqref{eq:geodesic motion x/T} if the spatio-temporal extent of the experiment is sufficiently small so that curvature terms are negligible.
In the limit $\velocity \to c$, the offsets given in \cref{eq:OC:offsets} reproduce those for light stated in \cref{eq:offsets:Stodolsky}. Similarly, the phase shift $\Delta \Eikonal_\textsc{oc}$ reproduces the optical result $\Delta \Eikonal_\textsc{s} = - g h l \omega / c$ provided one replaces the relativistic particle momentum $m v / \sqrt{1 - v^2/c^2}$ with the optical analog $\hslash k = \hslash \omega / c$.

\begin{figure}[t]
	\centering

	\subfloat[]{%
		\includegraphics[page=9]{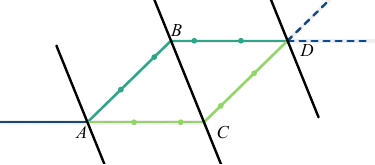}
		\label{fig:COW-gravity}
	}

	\subfloat[]{%
		\includegraphics[page=10]{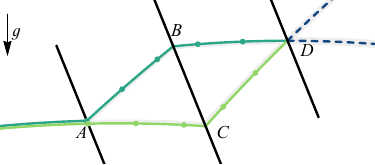}
		\label{fig:COW+gravity}
	}

	\caption{
		Schematic representation of the neutron interferometer described by Colella, Overhauser, and Werner in Ref.~\cite{1975PhRvL..34.1472C}.
		In the absence of gravity (a), the rays form a rhombus similar to that in the setup proposed by Overhauser and Colella, see \cref{fig:OC-gravity}, but in the presence of gravity (b) the free-fall trajectories differ from those shown in \cref{fig:OC+gravity}.
		As a consequence, the minimal beam width and the offsets at the first beam splitter (indicated by the gray band and the dots along the rays) are reduced.
		The overall phase shift, however, is the same in both setups.
	}
	\label{fig:COW}
\end{figure}

In their 1975 paper, Colella, Overhauser, and Werner (\COW) reported on the experimental observation of gravitationally induced phase shifts in single neutrons using a setup whose geometry differs from that originally proposed by Overhauser and Colella \cite{1975PhRvL..34.1472C}, see \cref{fig:COW} for a simplified schematic.
Compared to the Overhauser–Colella setup shown in \cref{fig:OC}, the incoming rays are sent towards one of the outer crystal slabs, rather than the middle one. As a consequence, the free-fall trajectories in the Overhauser–Colella setup cross, but those in the \COW\ setup do not, cf.~\cref{fig:OC+gravity,fig:COW+gravity}.
Similarly to the Overhauser–Colella setup, the rays interfering at the output must have reached the first crystal slab at different times and at different locations:
\begin{subequations}
	\label{eq:COW:offsets}
	\begin{align}
		\Delta t_\textsc{cow}
			&= - \frac{\gravA h l}{\velocity^3} \left[ \tan^2 \theta + 1 - 2 \frac{\velocity^2}{c^2} \right]\,,
		\\
		\Delta x_\textsc{cow}
			&= - \frac{\gravA h l}{\velocity^2} \tan^2 \theta\,,
		\\
		\Delta z_\textsc{cow}
			&= - \frac{\gravA h l}{\velocity^2} \tan \theta\,,
	\end{align}
\end{subequations}
cf.~\cref{eq:OC:offsets}.
The overall phase shift, however, is identical to that of the previous setup:
\begin{align}
	\label{eq:phase:COW}
	\begin{split}
		\Delta \Eikonal_\textsc{cow}
			&= - \frac{\gravA h l}{\hslash \velocity^2} \frac{m \velocity}{\sqrt{1 - \velocity^2/c^2}}
			\\&
			= - \frac{\gravA h l m}{\hslash \velocity} \left[1 + O(\velocity^2/c^2)\right]\,.
	\end{split}
\end{align}
Similarly to the Overhauser–Colella phase shift, there are multiple derivations for \cref{eq:phase:COW} based on various approximations, differing, in particular, in whether the rays are taken to be straight lines or free-fall parabolas, cf., e.g., Refs.~\cite{1986PhLB..182..211B,2025RvMP...97a5003B}.
\Cref{eq:phase:COW} can be generalized to account for the non-zero thickness of the crystal slabs \cite{1975PhRvL..34.1472C,1984PhRvD..30.1214B,1986PhyBC.137..260H} but the detailed description of such non-gravitational effects is beyond the scope of this paper.

Contrary to the optical experiments described in \cref{s:Tanaka,s:Stodolsky}, the parameters entering \cref{eq:OC:phase shift,eq:phase:COW} are not independent since the diffraction of the rays must satisfy the Bragg–Laue condition
$\bs n \cdot (\bs p_\text{out} - \bs p_\text{in}) = \hslash\, \bs n \cdot \bs k_\text{lattice}$,
where $\bs k_\text{lattice}$ is a reciprocal lattice vector and $\bs n$ is a collinear unit vector\footnote{Specifically, if $d_\text{lattice}$ denotes the lattice constant along $\bs n$, $\bs k_\text{lattice}$ must be an integer multiple of $(2 \pi / d_\text{lattice}) \bs n$, with the absolute value of this integer coefficient being commonly referred to as the order of the diffraction process (the diffraction in the \COW\ experiment was of first order). In practice, reflection also occurs for momenta that do not satisfy the diffraction condition \emph{exactly}, with the finite range of admissible $\bs p_\text{in}$ being known as the Darwin plateau.}.
Using this notation, for $\velocity \ll c$, \cref{eq:phase:COW} can be written in the form
\begin{align}
	\label{eq:phase:COW lattice tof}
	\Delta \Eikonal_\textsc{cow}
		&= - \bs \gravA \cdot \bs k_\text{lattice} \, T_*^2\,,
\end{align}
where $T_* = \length/\velocity$ is the time of uninterrupted free fall (up to negligible corrections of higher order), see also Ref.~\cite[App.~A]{2011CQGra..28n5017W}. The observation that the parameters $h$, $\length$, and $m$ in \cref{eq:OC:phase shift,eq:phase:COW} are constrained to a two-dimensional parameter space resolves the tension between the statement that “the phase shifts in atomic beam and neutron interferometry in general do depend on the mass” \cite{1996GReGr..28.1043L} on the one hand, and that “the phase shift is independent of the mass of the atoms (neutrons) or the associated Compton frequency” \cite{2011CQGra..28n5017W} on the other.

Similarly to the optical experiments described above, the relative size of the shift-induced phase difference can be estimated by the ratio $2 (\cos \vartheta) \velocity \Omega /\gravA$. For the \COW\ experiment, for example, this evaluates to $\sim \num{4e-2} \cos \vartheta$. The reason for this suppression of shift-induced effects compared to optical setups lies in the velocity ratio $\velocity /c \ll 1$. Hence, decreasing the particle’s velocity reduces the relative error produced by neglecting the shift-induced phase.

As shown in \cref{eq:eikonal:light,eq:eikonal:matter}, the gravitationally induced phase shift in neutrons is larger than that in photons by a factor that is given  by the ratio of the optical wavelength to the neutron’s de~Broglie wavelength. As a result, gravitational effects can be measured using interferometers with arm lengths of only a few centimeters \cite{1997PhRvA..56.1767L}, in contrast to the kilometer-scale interferometers required for optical experiments, as discussed above. However, current results from \COW\ experiments on neutron interferometry can also be derived using a purely Newtonian gravitational description, and as such, do not provide further insight into the combined framework of quantum theory and general relativity. In order to measure effects indicative of general relativistic theory such as space-time curvature or Lense–Thirring precession, a much larger interaction area would be required, since the corrections from higher-order gravitational effects are on the order of \num{e-9} of the \COW\ phase \cite{1984PhRvD..30.1615A}. Recent advancements in neutron interferometry include the fabrication of large, defect-free silicon crystals or use of other materials for neutron diffraction \cite{2024ApPhL.124g1901H,2024PhRvR...6d3079H}, very cold neutron sources \cite{2023JNR...24bM} and improvements in environmental shielding and detection \cite{2015AHEP...2015.687480P}. While these are increasing the sensitivity of neutron interferometry, it is still limited by small interaction areas, pushing gravitational research with massive particles towards atom fountain experiments that allow for a much larger interaction length (see \cref{s:fountains}). Alternative proposals include focusing on using the neutron's spin precession in a magnetic field as a “clock” to track the proper time of the neutron’s evolution, and thus allowing measurement of the proper time difference across the gravitational field \cite{2011NatCo...2..505Z}. However, these effects would be minuscule ($\sim \num{e-10}$ of the \COW\ phase for a magnetic field of \SI{10}{\tesla}), and are thus out of reach for current technologies. An in-depth discussion on possible neutron experiments involving gravity is provided in Ref.~\cite[Chap.~8]{2015_Rauch_Werner}.

\subsection{Atomic fountains}
\label{s:fountains}
Whereas the interferometers described above enclose non-zero areas, in atomic-fountain experiments the particle motion is commonly approximated as being restricted to one spatial dimension \cite{2001Metro..38...25P,2021AmJPh..89..324O}.
More generally, if motion takes place in the vicinity of a spatial curve, one can locally erect spatial Fermi coordinates $(x, y, z) \equiv (\t x{^\alpha}, z)$, relative to which the spatial metric $\t\metric{_i_j}$ takes a particularly simple form \cite{2025PhRvR...7a3162M}. If the reference curve is a spatial geodesic (as is relevant, for example, for radial motion in spherically symmetric geometries), one has
\begin{align}
	\begin{split}
		\metric
			&= [1 - \t\riemann{_z_\alpha_z_\beta} \t x{^\alpha} \t x{^\beta }] (\d z)^2
			\\&\quad
			+ [\t\delta{_\alpha_\beta} - \third \t\riemann{_\alpha_\gamma_\beta_\delta} \t x{^\gamma} \t x{^\delta}] \t{\d x}{^\alpha} \t{\d x}{^\beta}
			\\&\quad
			+ O[(\t\delta{_\alpha_\beta} \t x{^\alpha} \t x{^\beta})^{3/2}]\,,
	\end{split}
\end{align}
where the components of the spatial curvature tensor $\t\riemann{_i_j_k_l}$ are evaluated on the reference curve (along which $\t x{^\alpha} = 0$).
If motion takes place only along the reference curve, the exact equations of motion \eqref{eq:geodesic motion x/T} in a static reference frame take the form
\begin{subequations}
\begin{align}
	\ddot z &= - (1 - \dot z^2 / c^2) \llapse'(z)\,,
	\\
	\dot t &= \exp[- \llapse(z) / c^2]\,,
\end{align}
\end{subequations}
where dots indicate derivatives with respect to the temporal parameter $T$.
Although these equations generally do not admit closed-form solutions, for practical applications it suffices to use short-time expansions of the form
\begin{subequations}
\label{eq:fountain trajectories perturbative}
\begin{align}
	\begin{split}
		t(T)
			&= t_0
			+ \frac{\e^{-\llapse(z_0)/c^2}}{\tilde\gamma_0}
			\bigg\{
				\tilde\gamma_0 T
				- \frac{(\tilde\gamma_0 T)^2}{2!} \frac{p_0}{m c} \frac{\llapse'_0}{c}
				\\&\quad
				- \frac{(\tilde\gamma_0 T)^3}{3!} \bigg[
					\frac{p_0^2}{m^2 c^2} \llapse''_0
					- \bigg( 1 + \frac{p_0^2}{m^2 c^2} \bigg) \frac{(\llapse'_0)^2}{c^2}
				\bigg]
				\\&\quad
				+ O(T^4)
			\bigg\}\,,
	\end{split}
	\\
	\begin{split}
		z(T)
			&= z_0
			+ \frac{p_0}{m} \tilde\gamma_0 T
			- \frac{(\tilde\gamma_0 T)^2}{2!} \llapse'_0
			\\&\quad
			- \frac{(\tilde\gamma_0 T)^3}{3!} \frac{p_0}{m} \bigg( \llapse''_0 + 2 \frac{(\llapse'_0)^2}{c^2}\bigg)
			+ O(T^4)
			\,,
	\end{split}
	\\
	\begin{split}
		\tau(T)
			&= \tau_0
			+ \tilde\gamma_0 T
			+ \frac{(\tilde\gamma_0 T)^2}{2!} \frac{p_0}{m c} \frac{\llapse'_0}{c}
			\\&\quad
			+ \frac{(\tilde\gamma_0 T)^3}{3!} \bigg[
				\frac{p_0^2}{m^2 c^2} \llapse''_0
				- \bigg( 1 - \frac{p_0^2}{m^2 c^2} \bigg) \frac{(\llapse'_0)^2}{c^2}
			\bigg]
			\\&\quad
			+ O(T^4)\,,
	\end{split}
\end{align}
\end{subequations}
where subscript zeros indicate initial values at $T = 0$, $p = m \dot z / \sqrt{1 - \dot z^2/c^2}$ denotes the momentum, and $\tilde\gamma = 1/\sqrt{1 + p^2 / m^2 c^2}$ is the reciprocal Lorentz factor.
For simplicity of exposition, the following analysis is carried out in the limit $c \to \infty$, in which case \cref{eq:fountain trajectories perturbative} reproduces the trajectories of Newtonian mechanics. Corrections beyond this limiting case are described in Appendix~\ref{s:atom fountains:FSL}.

Whereas \cref{eq:eikonal:light,eq:eikonal:matter} suffice for computing the phase shifts in the experiments described above, this is no longer the case here since the atom-light interactions produce non-trivial phase shifts.
Specifically, the interaction of the atoms with $\pi$- and $\pi/2$-pulses produces additional phase shifts of the form
\begin{align}
	\label{eq:phases Bragg Raman}
	\Delta\Eikonal_*
		= \sum_i \left[
			\t*n{^\I_i}\, \Eikonal_\EM(\t*t{^\I_i}, \t*z{^\I_i}) - \t*n{^\II_i}\, \Eikonal_\EM(\t*t{^\II_i}, \t*z{^\II_i})
		\right]\,.
\end{align}
Here, $i$ extends over all atom-light interactions at the coordinates $(\t*t{^\I_i}, \t*z{^\I_i})$ in the upper arm and at $(\t*t{^\II_i}, \t*z{^\II_i})$ in the lower arm, $\Eikonal_\EM$ denotes the phase of the electromagnetic field, and the integers $\t*n{^\I_i}$ and $\t*n{^\II_i}$ depend on the details of the interaction \cite{2001Metro..38...25P,2013NJPh...15b3009A}:
For Bragg transitions, the absolute values of $\t*n{^\I_i}$ and $\t*n{^\II_i}$ indicate the order of the transition, whereas $|\t*n{^\I_i}| = |\t*n{^\II_i}| = 1$ for Raman transitions, and the signs of $\t*n{^\I_i}$ and $\t*n{^\II_i}$ indicate whether the atom’s momentum along the gradient of $\Eikonal_\EM$ increases or decreases in the process of the interaction.
A heuristic derivation of \cref{eq:phases Bragg Raman} can be found in Ref.~\cite[Sect.~3.2]{2009aosp.conf..411H}; more detailed derivations for Bragg and Raman transitions are provided in Ref.~\cite{1981PhRvA..23.1290B} and Ref.~\cite{1992PhRvA..45..342M}, respectively.
If Raman transitions are used, additional phase shifts arise if the two interfering rays spend unequal times in the various internal energy states. However, since the experiments considered here have equal time intervals between the light-matter interactions, these additional phase contributions vanish \cite{2011CQGra..28n5017W}.

\begin{figure}[b]
	\centering

	\subfloat[]{%
		\includegraphics[page=1]{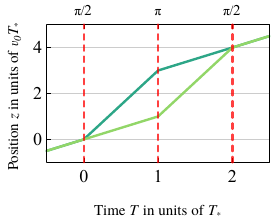}%
		\label{fig:fountain inertial}%
	}
	\hfill
	\subfloat[]{%
		\includegraphics[page=2]{fountains.pdf}%
		\label{fig:fountain homogeneous}%
	}
	\hfill
	\subfloat[]{%
		\includegraphics[page=3]{fountains.pdf}%
		\label{fig:fountain inhomogeneous}%
	}

	\caption{
		Schematic space-time diagrams of atomic fountain interferometers in an inertial system (a), in a uniform gravitational field (b), and in an inhomogeneous gravitational field (c).
		Atoms are launched into a vacuum tube and manipulated using a sequence of laser pulses.
		In the absence of gravity gradients, both rays take equal amounts of proper time regardless of the vertical acceleration $\gravA$.
		The strength and form of the gravitational field determines the position of the light-matter interactions and thus also the phase shifts acquired in the course of these interaction events.
		If the relative recoil velocities are in the ratio $1:-2:1$, the rays interfering at the output are separated at the input by a height difference that depends on the gravity gradient (b) and (c).
	}
	\label{fig:fountains}
\end{figure}

\paragraph{Gravimetry}
In their 1991 paper, Kasevich and Chu reported on the first atomic-fountain experiment capable of functioning as a gravimeter \cite{1991PhRvL..67..181K}. While this experiment used stimulated Raman transitions (see Ref.~\cite{1992ApPhB..54..321K} for a detailed model of the experiment), in 2013, Altin \etal.\ reported on a refined experiment using Bragg transitions instead \cite{2013NJPh...15b3009A}.
Simplified space-time diagrams of these experiments are shown in \cref{fig:fountain inertial,fig:fountain homogeneous}: a $\half\pi$-pulse at $T = 0$ acts as a beam splitter to produce atomic wave packets that propagate upwards with different velocities $\velocity_0$ and $\velocity_0 + \Delta\velocity_\text{r}$, respectively, where $\Delta\velocity_\text{r} = n \hslash k_\EM /m$ is the recoil velocity (in which $n \equiv \t*n{_\I^1}$ denotes the order of the diffraction process). After the “interrogation time” $T_*$, the wave packets are subject to a $\pi$-pulse that changes the velocities by $\pm n \hslash k_\EM /m$. The wave packets meet at $T = 2T_*$, where a final $\half\pi$-pulse brings them to equal final velocities in each output arm, after which interference is observed.
Integrating \cref{eq:phase evolution:matter} along the trajectories given by \cref{eq:fountain trajectories perturbative} shows that $\Delta \Eikonal_\lapse$ vanishes if gravity gradients and $\velocity^2/c^2$-corrections are negligible. This is in  agreement with calculations based on the Schrödinger equation for a particle in a linear gravitational potential \cite{2012Giulini}. Assuming low velocities, i.e., $\velocity \ll c$, and an electromagnetic phase of the form $\Eikonal_\EM = \omega_\EM (z/c - t)$, one obtains
\begin{subequations}
	\begin{align}
		\Delta \Eikonal_\lapse &= 0\,,
		\\
		\Delta \Eikonal_*
			&= - \gravA m \Delta\velocity_\text{r} T_*^2 /\hslash^2
			 = - n \,\bs \gravA \cdot \bs k_\EM T_*^2\,,
	\end{align}
\end{subequations}
where $\gravA = \d \llapse / \d z$ is the vertical gravitational acceleration.
Hence, the overall phase shift is structurally identical to that of \cref{eq:phase:COW lattice tof}.
The interpretation of this result was debated in the literature \cite{2010Natur.463..926M,2010Natur.467E...1W,2010Natur.467E...2M,2011CQGra..28n5017W,2011CQGra..28n5018S,2012Giulini,2012CQGra..29d8002W,2013NJPh...15a3007S}.
Specifically, in view of \cref{eq:phase evolution:matter}, Müller \etal.\ interpreted the phase shift as a measure of proper time tracked by a “Compton clock” \cite{2010Natur.463..926M}. However, Sinha and Samuel \cite{2011CQGra..28n5018S} as well as Giulini \cite{2012Giulini} argued that \cref{eq:phase evolution:matter} does not describe any observable periodic changes in internal degrees of freedom. Moreover, Wolf \etal.\ showed that both interfering rays have equal proper times \cite{2010Natur.467E...1W,2011CQGra..28n5017W}. This last observation is consistent with $\Delta \Eikonal_\lapse = 0$ derived here.

In a subsequent experiment, Asenbaum \etal.~\cite{2020PhRvL.125s1101A} compared high-precision phase measurements from an atom cloud of \textsuperscript{87}Rb to that of \textsuperscript{85}Rb in order to measure differential gravitational accelerations between the two isotopes. The results showed no deviations from universality of free fall to the level of $\num{e-12} \gravA_\text{std}$, where $\gravA_\text{std}$ denotes the standard acceleration of gravity.
The atom clouds ($\sim \num{e5}$ atoms per cloud) underwent evaporative cooling before being launched from the atom trap to a height of \SI{8.6}{\meter}. With a total free-fall time of $2T_* = \SI{1910}{\milli\second}$, the experiment had a single-shot acceleration sensitivity of $\num{1.4e-11} \gravA_\text{std}$. Experimental challenges to reach this sensitivity involve an \textsc{ac}-Stark shift generated by the atom launch optics, which can be reduced by improved laser systems; differential magnetic gradients across the drop tower, mitigated significantly by generating a uniform magnetic field across the experiment with a solenoid; and imaging errors, which are addressed by calibration of the atom imaging optics.

\paragraph{Gradiometry}
\label{s:fountains:gravimeters}
Using multiple gravimeters at different heights, it is possible to measure gravity gradients $\gamma = \d \gravA / \d z$ \cite{1998PhRvL..81..971S,2001Metro..38...25P,2002PhRvA..65c3608M,2007Sci...315...74F,2012ApPhL.101k4106S,2017PhRvL.118r3602A} and even second spatial derivatives of the gravitational acceleration \cite{2015PhRvL.114a3001R}.
Additionally, it is possible to extend the free-fall heights of atomic-fountain experiments so that higher-order terms in \cref{eq:fountain trajectories perturbative} become relevant.
Specifically, if $T^3$-terms are retained in the perturbative expansion and if the same electromagnetic pulse sequence is used as for the gravimeters described above, the rays interfering at the output must have a vertical offset $\Delta z_0 = \gamma \Delta\velocity_\text{r} T_*^3 + O(T_*^4)$ at the input \cite{2001CRAS..2..509B}, see \cref{fig:fountain inhomogeneous}.
In this case (assuming, as above, $\velocity \ll c$), the total phase shift takes the form $\Delta\Eikonal = \Delta\Eikonal_\lapse + \Delta\Eikonal_*$ with
\begin{subequations}
\label{eq:gradiometer phase}
\begin{align}
	\Delta \Eikonal_\lapse
		&= \half \gamma m \Delta\velocity_\text{r}^2 T_*^3 / \hslash + O(T_*^4)
		\,,
	\\
	\begin{split}
		\Delta \Eikonal_*
			&=
			- \gravA  m \Delta\velocity_\text{r} T_*^2 / \hslash
			\\&\quad
			- \gamma m \Delta\velocity_\text{r} ( \velocity_0 + \Delta\velocity_\text{r}) T_*^3/\hslash
			+ O(T_*^4)\,,
	\end{split}
\end{align}
\end{subequations}
in agreement with calculations based on the Schrödinger equation with a Newtonian potential \cite{1999PhLA..251..241W}.
Alternatively, the frequencies of the electromagnetic pulses can be adjusted (depending on the gravity gradient $\gamma$) to make the wave packets overlap at both the input and the output. In this case, the total phase shift takes the same value as given here \cite{2011CQGra..28n5017W}.
Such a gravity-gradient-induced phase shift was measured by Overstreet \etal.\ in 2022 \cite{2022Sci...375..226O}.

\Cref{eq:gradiometer phase} also shows that gravity gradients can introduce systematic errors in measurements of $\gravA$. For high-precision gravimetry, additional gravity-gradient terms of order $T_*^4$ must also be accounted for, see Appendix~\ref{s:atom fountains:T4}.

Atomic fountain experiments of this kind are candidates for many future gravitational experiments. One such proposal suggests using atomic fountains to detect gravitational waves (\GW), allowing measurement of signals out of reach for current optical detection systems \cite{2009PhLB..678...37D,2023arXiv230400614A,2025arXiv250911867B}. These include both terrestrial and space-based detection systems, consisting of multiple atom interferometers measuring variations in the timing of the laser pulses caused by a \GW\ passing through the interferometer. Other space-based atom interferometry experiments have been proposed for further tests of equivalence principles and for dark matter searches \cite{2019EPJD...73..228T}. Recent advancements in atom lattice traps are allowing the possibility of gravitational measurements with trapped atoms \cite{2024Natur.631..515P}. These experiments have the advantage of extended interaction times of tens of seconds. However, a theoretical description of these guided atom evolution is beyond the scope of this paper, and is one avenue for future expansion of the framework described here.

\section{Discussion}

In this paper, we developed a unified theoretical framework for describing quantum interference in both massless and massive systems in the presence of external stationary gravitational fields. Compared to perturbative weak-field expansions, our geometric approach offers a simpler and more transparent description, reducing interpretational ambiguities:

\begin{itemize}
	\item Our geometric description does not rely on any flat reference geometry and thus eliminates complications in modeling experimental setups using charts for which coordinate differences and lengths computed using a flat reference metric $\t\eta{_\mu_\nu}$ differ from physical lengths computed using the full metric $\t\Metric{_\mu_\nu}$, cf., e.g., Ref.~\cite{2018CQGra..35x4001B}.
	\item This approach makes it evident that details of the spatial metric tensor do not matter for the description of the experiments considered here. In particular, none of the experiments discussed here are sensitive to the parameter $\gamma_\PPN$ arising in the parameterized post-Newtonian (\PPN) framework. Previous derivations of this statement were restricted to optical interferometry and required significantly more extensive computations \cite{2015PhRvD..91f4041B,2022PhRvA.106f3511M}.
	\item Similarly, the general equations describing the phase evolution of massive and massless particles do not involve the space-time curvature tensor. Hence, curvature components can only be inferred by indirect means, e.g., by varying external parameters such as the spatial extent of the interferometer. In this context, we derived an exact formula that relates specific components of the space-time curvature tensor to the gravitational acceleration and its gradient, see \cref{eq:Riem:acceleration gradient}. This provides a rigorous basis for the commonly used heuristic of interpreting gravity gradients as an indicator for curvature.
	\item Moreover, our computation of lapse-induced phases in photons does not rely on heuristic methods such as effective refractive indices arising from a gravitational potential, nor does it require assigning to photons an effective gravitational mass.
\end{itemize}

We derived general formulas for the phase evolution of different quantum systems and used them to analyze various experimental configurations.
For experiments with photons, we showed that free-space and fiber-based interferometers yield essentially the same detectable lapse-induced phase shift. In both scenarios, the use of entangled quantum states provides new insights into the gravitational coupling as the resulting detection probabilities cannot be mimicked by classical field intensities. In experiments where gradients of the lapse-induced phase shifts are measurable, such setups enable the direct observation of specific components of the Riemann curvature tensor with non-classical probes. Our geometric approach also yields predictions for phase shifts in experiments that probe gravity using unguided matter-waves instead of photons. The phase shifts we derived for Mach–Zehnder, Bonse–Hart, and atomic fountain configurations not only agree with previous results obtained assuming either Newtonian gravity or linearized  Einsteinian gravity, but our method also extends these results to arbitrary stationary gravitational fields. The interpretation of the phase shifts arising for both the optical and matter-wave experiments follows naturally from our description. While gravitational phase shifts in optical interferometers arise directly from the lapse function, the situation is more subtle for matter-wave interferometers. In matter-wave gravimeters, the lapse-induced phase shift vanishes and the observed phase originates from the interaction with light pulses at different locations that depend on the gravitational acceleration. However, when gravity gradients are taken into account, lapse-induced phase shifts re-emerge in addition to the phase from the light-matter interaction and are directly proportional to the gravity gradient. Our methods can be extended to include internal degrees of freedom and also guided motion of massive particles.

\section*{Acknowledgements}
We thank Maximilian Ofner and Francesco Giovinetti for useful discussions.
This research is funded by the European Union (\ERC, \GRAVITES, project no.~101071779). Views and opinions expressed are however those of the authors only and do not necessarily reflect those of the European Union or the European Research Council Executive Agency. Neither the European Union nor the granting authority can be held responsible for them.
TM is supported by the Vienna Doctoral School in Physics (\VDSP).

\appendix

\section{Phase shifts in linearized gravity}
\label{s:linearized gravity Stodolsky}

In linearized gravity, the metric tensor is decomposed as $\t\Metric{_\mu_\nu} = \t \eta{_\mu_\nu} + \t\varepsilon{_\mu_\nu}$, where $\t \eta{_\mu_\nu}$ is the flat Minkowskian metric, and squares and higher powers of the metric perturbation $\t\varepsilon{_\mu_\nu}$ are assumed to be negligible \cite[Chap.~4.4]{1984_Wald}.
In this setting, versions of \cref{eq:phase:Stodolsky,eq:phase:COW} can be derived by integrating \cref{eq:phase evolution general} along curves other than the actual rays:
Denoting by $\t*\WaveC{^0_\mu}$ the “unperturbed” wave-covector satisfying $\t\eta{^\mu^\nu} \t*\WaveC{^0_\mu} \t*\WaveC{^0_\nu} = - m^2 c^2 / \hslash^2$ [with $m = 0$ for light], the eikonal equation implies that the “perturbed” wave-covector in linearized gravity can be written as
\begin{align}
	\t\WaveC{_\mu}
		= \t*\WaveC{^0_\mu}
		+ \half \t\varepsilon{_\mu_\nu} \t*\WaveV{_0^\nu}
		+ \t \kappa{_\mu}\,,
\end{align}
where $\t*\WaveV{_0^\mu} = \t\eta{^\mu^\nu} \t*\WaveC{^0_\nu}$ is the “unperturbed” wave-vector and $\t \kappa{_\mu}$ is any one-form satisfying $\t*\WaveV{_0^\mu} \t\kappa{_\mu} = 0$.
For any two points $p_1$ and $p_2$ lying along an integral curve $\Gamma$ of $\t*\WaveV{_0^\mu}$, \cref{eq:phase evolution general} yields
\begin{align}
	\label{eq:phase:Stodolskys formula}
	\Eikonal(p_2)
		= \Eikonal(p_1)
		+ \int_\Gamma [\t*\WaveC{^0_\mu} + \half \t\varepsilon{_\mu_\nu} \t*\WaveV{_0^\nu}] \t{\d x}{^\mu}\,,
\end{align}
in which $\t \kappa{_\mu}$ does not contribute because of its orthogonality to $\t*\WaveV{_0^\mu}$ \cite[Sect.~8]{1979GReGr..11..391S}.
When using this equation, care must be taken to use appropriate curves along which the integrals are to be evaluated.
In particular, these curves \emph{cannot} be prescribed independently of $\t\varepsilon{_\mu_\nu}$.
This can be seen, for example, by studying the transformation of the integral arising in \cref{eq:phase:Stodolskys formula} under linearized diffeomorphisms, under which $\t\varepsilon{_\mu_\nu}$ changes as $\text{\textdelta} \t\varepsilon{_\mu_\nu} = \t\p{_\mu} \t\lambda{_\nu} + \t\p{_\nu} \t\lambda{_\mu}$. If the curve $\Gamma$ were invariant under such gauge transformations, then the integral $I_\Gamma = \int_\Gamma [\t*\WaveC{^0_\mu} + \half \t\varepsilon{_\mu_\nu} \t*\WaveV{_0^\nu}] \t{\d x}{^\mu}$ would transforms as
\begin{align}
	\text{\textdelta} I_\Gamma
		= (\t\WaveV{^\mu} \t\lambda{_\mu})(p_2) - (\t\WaveV{^\mu} \t\lambda{_\mu})(p_1)\,,
\end{align}
where $p_1$ and $p_2$ are the initial and final endpoints of $\Gamma$, respectively.
This shows that gauge-invariant predictions for phase shifts can be computed using \cref{eq:phase:Stodolskys formula} only if the coordinate expressions for the curves involved are gauge-\emph{dependent}.

To illustrate this point, consider \cref{eq:phase:Stodolskys formula} applied to null rays with (angular) coordinate frequency $\omega$ propagating along straight coordinate lines joining the points
\begin{subequations}
	\label{eq:Stodolsky points}
	\begin{align}
		(\t*x{_A^i}) &= (0, 0, 0)\,,
		&
		(\t*x{_B^i}) &= (0, 0, h)\,,
		\\
		(\t*x{_C^i}) &= (\length, 0, 0)\,,
		&
		(\t*x{_D^i}) &= (\length, 0, h)\,.
	\end{align}
\end{subequations}
If $\t\metric{_i_j} = \t\delta{_i_j}$, the parameters $h$ and $\length$ accurately represent the physical arm lengths of the interferometer.
Integrating \eqref{eq:phase:Stodolskys formula} in the linear potential $\llapse = \gravA z$, one obtains the same result as in \cref{eq:phase:Stodolsky}.
If, instead, the metric is given by $\t\metric{_i_j} = (1 + 2 \gravA z / c^2)\t\delta{_i_j}$, which corresponds to a post-Newtonian metric with a linear potential \cite[§18.4]{1973_Misner_Throne_Wheeler}, then the geodesic distances of the “horizontal” segments are unequal:
\begin{align}
	\length_{AC} &\approx \length\,,
	&
	\length_{BD} &\approx (1 + \gravA h /c^2) \length\,.
\end{align}
This mismatch causes an additional phase shift $\Delta \Psi_\text{mismatch} = - \omega (\length_{AC} - \length_{BD}) / c \approx \omega \gravA h \length /c^3$, so that the overall phase shift is twice as large as that in \cref{eq:phase:Stodolsky}:
\begin{align}
	\label{eq:phase:Stodolsky imbalanced}
	\Delta \Psi_\text{imbalanced}
		&\approx - 2 \gravA h \length \omega /c^3\,.
\end{align}
Identical coordinate expressions for the spatial curves in \cref{eq:phase:Stodolskys formula} thus yield different phase shifts depending on the explicit form of $\t\varepsilon{_\mu_\nu}$, despite the fact that the two fields are related by a gauge transformation, namely that generated by
\begin{align}
	\t\lambda{_x}
		&= \gravA x z\,,
	&
	\t\lambda{_y}
		&= \gravA y z\,,
	&
	\t\lambda{_z}
		&= \half \gravA (z^2 - x^2 - y^2)\,.
\end{align}
This analysis also shows that experimental tests of \cref{eq:phase:Stodolsky} require precise control over the interferometer’s arm lengths, as relative imbalances of $\gravA h /c^2 \approx \num{e-16} (h / \si{\meter})$ may double or cancel the phase shift induced by the gravitational redshift.

\section{FSL corrections in atomic interferometers}
\label{s:atom fountains:FSL}

Because of the finiteness of the speed of light (\FSL), the optical pulses in an atomic fountain must be “chirped” to account for frequency shifts between the stationary reference frame and the atoms \cite{2001Metro..38...25P}, and the $\pi$-transitions in the two interferometer arms are not simultaneous \cite{2016PhRvA..94a3612T,2017PhRvD..95b4002T,2017PhRvA..96b3604T,2025AVSQS...7c4404N}.
The influence of these \FSL\ effects on the interferometric phase can be computed efficiently using the “retarded” time $\Tret(t, z)$ that measures at what proper time a light signal at zero altitude must be emitted in order for it to reach the point $(t, z)$. This function satisfies the eikonal equation
\begin{align}
	\t\Metric{^\mu^\nu} (\t\p{_\mu} \Tret) (\t\p{_\nu} \Tret) = 0
\end{align}
with boundary conditions $\Tret(t, z = 0) = t\, \e^{\llapse(0)/c^2}$ and can be computed perturbatively in $z$:
\begin{align}
	\begin{split}
		\Tret(t, z)
			&= t\, \e^{\llapse(0)/c^2}
			- \frac{z}{c}
			+ \frac{z^2}{2!} \frac{\llapse'(0)}{c^3}
			\\&\quad
			+ \frac{z^3}{3!} \left(
				\frac{\llapse''(0)}{c^3}
				- \frac{\llapse'(0)^2}{c^5}
			\right)
			+ O(z^4)\,.
	\end{split}
\end{align}
Since the electromagnetic phase is constant along null rays, the eikonal is of the form $\Eikonal_\EM(t, z) = f(\Tret(t, z))$, where $f$ describes the waveform as a function of proper time at $z = 0$, and hence $\d f / \d \Tret$ is the (negative) frequency emitted at $z = 0$ that is continuously adjusted to be resonant for the internal transition of the moving atoms.
The condition that the various light-matter interactions arise from light rays emitted at predetermined values of $\Tret$ can be implemented by parametrizing the world-lines by $\Tret$ instead of $T$.
Explicitly, the relation between the free-fall time $T$ and differences in retarded time $\Tret$ is given by
\begin{align}
	\begin{split}
		T &=
			\Delta\Tret \bigg(
				1
				+ \frac{p_0}{m c}
				+ \frac{p_0^2}{m^2 c^2}
				+ \frac{\llapse(z_0) - \llapse(0)}{c^2}
			\bigg)
			\\&\quad
			- \frac{(\Delta\Tret)^2}{2!} \bigg(
				\frac{\llapse'(z_0)}{c}
				+ \frac{p_0}{m c} \frac{\llapse'(z_0)}{c}
			\bigg)
			\\&\quad
			- \frac{(\Delta\Tret)^3}{3!} \bigg(
				\frac{p_0}{m c} \llapse''(z_0)
				+ \frac{p_0^2}{m^2 c^2} \llapse''(z_0)
				\\&\qquad\qquad
				- 2 \frac{\llapse'(z_0)^2}{c^2}
			\bigg)
			+ O((\Delta\Tret)^4)
			+ O(c^{-3})\,.
			\hspace{-5em}
	\end{split}
\end{align}
Extending the analysis in \cref{s:fountains} to account for these corrections leads to a phase shift of the form
\begin{align}
	\begin{split}
		\Delta \Eikonal_\lapse
			&= - \frac{\hslash k^2}{2 m c} \llapse'_0 T_*^2
			\\&\quad
			- \left(1 + \frac{4 p_0}{m c} + \frac{3 \hslash k}{
				m c} \right) \frac{\hslash k^2}{2 m} \llapse''_0 T_*^3
			\\&\quad
			+ O(T_*^4) + O(c^{-3})
	\end{split}
	\\
	\begin{split}
		\Delta \Eikonal_*
			&= - \left(1 + \frac{p_0}{m c}\right) k \llapse'_0 T_*^2
			\\&\quad
			- \left(1 + \frac{2 p_0}{m c}\right) \frac{k p_0}{m} \llapse''_0 T_*^3
			\\&\quad
			+ \frac{k}{c} (\llapse'_0)^2 T_*^3
			+ O(T_*^4) + O(c^{-3})
	\end{split}
\end{align}
In isolation, the results for $\Delta \Eikonal_\lapse$ and $\Delta \Eikonal_*$ do not converge to those given in \cref{eq:gradiometer phase}. This is because even though the world-lines with $c < \infty$ converge to the Newtonian trajectories as $c \to \infty$, the eikonal along them evolves as $\dd \Eikonal = - m c^2 \dd \tau / \hbar$, whose $c^2$-dependence makes the result depend on higher-order corrections to the trajectories. However, the observable phase shift $\Delta \Eikonal =  \Delta \Eikonal_\lapse + \Delta \Eikonal_*$ converges to that of \cref{eq:gradiometer phase} as $c \to \infty$, thus leading to unambiguous predictions.

More comprehensive models can be obtained by taking into account the finite interaction times of the atoms with the electromagnetic pulses. This leads to further $1/c$-contributions to the total phase, see, e.g., Ref.~\cite{2025AVSQS...7c4404N}.

\section{\texorpdfstring{$\boldsymbol{T_*^4}$}{T4}-Corrections in atomic interferometers}
\label{s:atom fountains:T4}

The perturbative expansion of \cref{s:fountains} can also be extended to arbitrary order in the time of flight.
For example, the fourth-order corrections to \cref{eq:fountain trajectories perturbative} are given by
\begin{align}
	\begin{split}
		t(T)_{[4]}
			&= - \frac{\e^{- \llapse(z_0)/c^2}}{\tilde\gamma_0} \frac{(\tilde\gamma_0 T)^4}{4!} \frac{p_0}{m c}
			\\&\quad\times
			\bigg[
				\frac{p_0^2}{m^2 c^2} c \llapse'''_0
				- \left(4 + \frac{3 p_0^2}{m^2 c^2}\right) \frac{\llapse''_0 \llapse'}{c}
				\\&\qquad
				+ \left(1 + \frac{p_0^2}{m^2 c^2}\right) \frac{(\llapse')^3}{c^3}
			\bigg]\,,
	\end{split}
	\\
	\begin{split}
		z(T)_{[4]}
			&= - \frac{(\tilde\gamma_0 T)^4}{4!} \bigg[
				\frac{p_0^2}{m^2} \llapse'''_0
				- \left(1 - \frac{6 p_0^2}{m^2 c^2}\right) \llapse''_0 \llapse'_0
				\\&\qquad
				- \left(2 - \frac{4 p_0^2}{m^2 c^2}\right) \frac{(\llapse'_0)^3}{c^2}\,,
			\bigg]\,,
	\end{split}
	\\
	\begin{split}
		\tau(T)_{[4]}
			&= \frac{(\tilde\gamma_0 T)^4}{4!} \frac{p_0}{m c}
			\\&\quad\times
			\bigg[
				\frac{p_0^2}{m^2 c^2} c \llapse'''_0
				- \left(5 - \frac{p_0^2}{m^2 c^2}\right) \frac{(\llapse'_0)^3}{c^3}
				\\&\qquad
				- \left(4 - \frac{3 p_0^2}{m^2 c^2}\right) \frac{\llapse''_0 \llapse_0}{c}
			\bigg]\,.
	\end{split}
\end{align}
In the limit $c \to \infty$, this results in the following correction to the phase shifts given in \cref{eq:gradiometer phase}:
\begin{align}
	\begin{split}
		\Delta\Eikonal_\lapse
			&= \frac{\hslash k^2}{2m} \llapse''_0 T_*^3
			+ \frac{7 \hslash k^2 \velocity_0}{12 m} \llapse'''_0 T_*^4
			\\&\quad
			+ \frac{\hslash^2 k^3}{3 m^2} \llapse'''_0 T_*^4
			+ O(T_*^5)\,,
	\end{split}
\end{align}
\begin{align}
	\begin{split}
		\Delta\Eikonal_* &%
			= - k \llapse'_0 T_*^2
			- k \velocity_0 \llapse''_0 T_*^3
			- \frac{\hslash k^2}{m} \llapse''_0 T_*^3
			\\&\quad
			+ \frac{7 k}{12} \llapse'_0 \llapse''_0 T_*^4
			- \frac{7 k \velocity_0^2}{12} \llapse'''_0 T_*^4
			\\&\quad
			- \frac{7 \hslash k^2 \velocity_0}{6 m} \llapse'''_0 T_*^4
			- \frac{\hslash^2 k^3}{2 m^2} \llapse'''_0 T_*^4
			+ O(T_*^5)\,,
			\hspace*{-5em}
	\end{split}
\end{align}
where $\velocity_0 = p_0 / m$.
For a constant gravity gradient, i.e., $\varphi''' = 0$, this reproduces the results of Refs.~\cite{1998Peters,1999PhLA..251..241W}.
Despite being of higher order in $T_*$, the contribution $\tfrac{7}{12} k \llapse'_0 \llapse''_0 T_*^4$ is numerically comparable to the third-order term $- k \velocity_0 \llapse''_0 T^3$ and thus needs to be taken into account in high-accuracy measurements \cite{2001Metro..38...25P}.
Writing the overall phase shift as $\Delta\Eikonal = - \mathfrak g k T_*^2$, the effective acceleration $\mathfrak g$ takes the form
\begin{align}
	\begin{split}
		\mathfrak g
			&= \llapse'_0
			+ \velocity_0 \llapse''_0 T_*
			+ \frac{\hslash k}{2m} \llapse''_0 T_*
			- \frac{7}{12} \llapse'_0 \llapse''_0 T_*^2
			\\&\quad
			+ \frac{7}{12} \velocity_0^2 \llapse'''_0 T_*^2
			+ \frac{7 \hslash k \velocity_0}{12 m} \llapse'''_0 T_*^2
			\\&\quad
			+ \frac{\hslash^2 k^2}{6 m^2} \llapse'''_0 T_*^2
			+ O(c^{-1})
			+ O(T_*^3)\,,
	\end{split}
\end{align}
which reproduces all significant contributions up to and including fourth order in $T_*$ listed in Table~I of Ref.~\cite{2009aosp.conf..411H}.
As was pointed out in Ref.~\cite{2016PhRvA..93b3617N}, the quantity $\mathfrak g$ differs from the gravitational acceleration of the atoms at the time $T_*$, see also Refs.~\cite{2018PhRvL.121l8903D,2018PhRvL.121l8904R,2020PhRvR...2a2036N}.

\bibliography{bibliography}
\end{document}